# Fast data augmentation for battery degradation prediction

Weihan Li[1,2*§], Harshvardhan Samsukha[1,2§], Bruis van Vlijmen[3,4], Lisen Yan[1,2], Samuel Greenbank[5], Simona Onori[3,6], Venkat Viswanathan[7]

[1] Center for Ageing, Reliability and Lifetime Prediction of Electrochemical and Power Electronic Systems (CARL), RWTH Aachen University, Campus-Boulevard 89, 52074 Aachen, Germany
[2] Institute for Power Electronics and Electrical Drives (ISEA), RWTH Aachen University, Campus-Boulevard 89, 52074 Aachen, Germany
[3] SLAC National Accelerator Laboratory, Menlo Park, California 94025, USA
[4] Department of Materials Science and Engineering, Stanford University, Stanford, California 94305, USA
[5] Department of Engineering Science, University of Oxford, Oxford, OX1 3PJ, UK
[6] Department of Energy Science and Engineering, Stanford University, Stanford, California 94305, USA
[7] Department of Aerospace Engineering, University of Michigan, Ann Arbor, Michigan 48109, USA

\* Correspondence: weihan.li@isea.rwth-aachen.de (W.L.)
§ Both authors contributed equally

## Abstract

Degradation prediction for lithium-ion batteries using data-driven methods requires high-quality aging data. However, generating such data, whether in the laboratory or the field, is time- and resource-intensive. Here, we propose a method for the synthetic generation of capacity fade curves based on limited battery tests or operation data without the need for invasive battery characterization, aiming to augment the datasets used by data-driven models for degradation prediction. We validate our method by evaluating the performance of both shallow and deep learning models using diverse datasets from laboratory and field applications. These datasets encompass various chemistries and realistic conditions, including cell-to-cell variations, measurement noise, varying charge-discharge conditions, and capacity recovery. Our results show that it is possible to reduce cell-testing efforts by at least 50% by substituting synthetic data into an existing dataset. This paper highlights the effectiveness of our synthetic data augmentation method in supplementing existing methodologies in battery health prognostics while dramatically reducing the expenditure of time and resources on battery aging experiments.

## Introduction

The application of lithium-ion batteries (LIBs) is widespread in many sectors of the economy due to their high gravimetric and volumetric energy density, long lifespan and decreasing production costs [1]. However, with repetitive charge-discharge cycles as well as storage, these electrochemical systems undergo degradation, which manifests itself in the form of capacity, energy, and power losses [2]. Therefore, accurate degradation prediction of LIBs has been an extensive area of research. Such insights can enable original equipment manufacturers (OEMs), such as electric vehicle (EV) makers, to assign calculated warranties to battery packs and ensure high pack performance and safety during their expected lifetime, while end-users and battery system operators can be informed of the date of pack replacement in advance. However, reliable and accurate degradation prediction remains challenging due to the nonlinear nature of such systems that stem from internal electrochemical reactions and intrinsic parameter variability across cells.

Several studies have proposed physics-based models [3–9] and semi/-empirical models [10–13] to characterize and parametrize the degradation mechanisms of LIBs. Although these models have demonstrated their predictive abilities, it remains challenging to capture nonlinear degradation trends and account for intrinsic manufacturing variabilities while sufficiently parametrizing a broad range of LIB degradation mechanisms is challenging. Moreover, not all aging mechanisms have been well understood and accurately modeled in the laboratory. In contrast, the availability of increasing amounts of field and laboratory data has paved the way for data-driven approaches [10]. Numerous studies have used both shallow machine learning [11–15] and deep learning [16–20] models to predict the degradation of LIBs. Rather than modeling the underlying degradation mechanisms in LIBs with predefined equations, these models derive relations directly from battery aging data. Therefore, they require a large quantity of data to train the model to capture degradation trends successfully. In general, the performance of data-driven methods is affected by the inadequate quantity of high-quality data.

The primary problems encountered in generating high-quality datasets for such models are insufficient numbers and types of aging scenarios and inadequate numbers of cells tested per aging scenario [21, 22]. Recent trends have shown rapid improvements in existing LIB compositions as well as the development of new LIB chemistries [23]. Thus, thoroughly examining the performance characteristics of newly developed LIB materials within a limited time before the application is of utmost priority. At the same time, developing the degradation prediction model with limited data and time is also important. In addition, cell testing in laboratories is resource-, time- and space-intensive [24], and there is an urgent need among research groups and OEMs to accelerate cell aging tests at lower costs.

One possible solution is the generation of synthetic data. Based on the available literature, methods to generate synthetic data can be categorized into electrochemical modeling, mechanistic modeling, and data-driven approaches. Electrochemical modeling, of which the pseudo-two-dimensional (P2D) model is the most widely used framework, computes the evolution of the internal states of the battery based on physical principles. Due to a large number of parameters and high computational complexity, it is challenging and time-consuming to use these models to generate a large synthetic dataset. Moreover, it is unclear how the physical parameters change in different aging scenarios, adding another layer of uncertainty. Secondly, the mechanistic modeling framework proposed by Dubarry parametrizes the loss of active material and lithium inventory to simulate electrode half-cell curves under different degradation pathways [25, 26]. Consequently, one could simulate full cell voltage curves and the capacity fade curves using minimal computational effort. Despite these favorable qualities, the mechanistic modeling approach can pose practical limitations as it requires the availability of reference half-cell voltage curve data. This prerequisite could require one to dismantle the cell and produce half cells [25], which is not available in most application cases. Additionally, replicating this method on different cell types is expected to be cost-intensive due to the need for half-cell voltage curve data of each different cell chemistry.

Data-driven methods have seen significant growth, particularly with advancements in Generative AI technologies like Generative Adversarial Networks (GANs) and Autoencoders [27, 28]. Jiang et al. [29] proposed an enhanced Conditional Variational Autoencoder to generate operational data with a specific focus on the charging process. Various machine learning models, including neural networks [30] and transformer-based deep learning models [31], have been developed to predict voltage discharge curves and estimate the current aging state. These studies underscore the importance of generating operational data for both charging and discharging processes, though further validation is needed for their application in lifetime prediction. Lin et al. [32] introduced a model combining polynomial functions with neural networks to generate synthetic capacity degradation curves. However, the need to manually partition the dataset into subgroups based on cycle numbers limits its scalability, making it more suited for smaller, less diverse datasets. Another method [33] used machine learning approaches to reconstruct missing battery data using a few correctly measured data points. However, such a reconstruction is only possible if the cells have been aged till their End of Life (EOL) under normal or accelerated tests. This method is more suited for recovering field data where cells have already been aged under the application load profile. A quick data augmentation method with a low computation burden, which doesn't need invasive half-cell characterization and material information to boost the data-driven aging prediction, is missing.

In this work, we focus on proposing a practical and efficient data generation method to enhance the performance of existing data-driven degradation prediction models. We generate synthetic capacity fade data using a function consisting of three parameters that can potentially model all possible electrochemical mechanisms of degradation in LIBs based on an existing seed dataset. A massive amount of synthetic degradation data can be generated through this method within a few seconds and with low computing costs. The generalizability of this simple method relaxes the experimental requirements of the seed dataset, allowing room for obtaining specific or extreme cell degradation scenarios for cells independent of their chemistries. Knowledge of complex battery aging mechanisms is also not necessary to generate the data. The method could serve to enrich sparse datasets - thereby reducing time and resource expenditure on cell aging tests. Or on the other hand, it generates data with additional variations in existing aging datasets to potentially increase the prediction performance of data-driven models. In the study, our method is validated by assessing the performance of both shallow machine learning and deep learning models, predicting the future capacity fade trajectory from early life data in the following scenarios: (a) seed dataset with different amounts of real cell data, (b) seed dataset with different amounts of real cell data + fixed amount of synthetic cell data, (c) partially replacing real cell data of seed dataset with synthetic cell data, (d) enriching sparse seed datasets with synthetic cell data. Our results show an improvement in the prediction performance of the data-driven models and a possibility of effort reduction for cell testing by at least 50%.

## Framework

The general workflow of the tasks in this paper is shown in Figure 1a. The initial step is obtaining seed data from aging tests and thorough data analysis to characterize the statistics of the seed dataset, followed by building models for the degradation prediction of unseen cells. The steps for synthetic data generation and its utilization are integrated after the data analysis step to build better prediction models. The method of synthetic data generation works by modifying the capacity fade curves of an existing battery degradation dataset, i.e., seed dataset, as shown in Figure 1b. The proposed method of creating synthetic degradation curves involves a series of steps. The initial capacity values and the final cycle numbers are identified for the capacity degradation curves within the seed dataset. These are used to define the parameters of the synthetic data generation function, namely *Offset*, *Slope* and *Elongation*, each representing a characteristic of the variation in the capacity degradation curves of LIBs.

These three parameters, *Offset*, *Slope* and *Elongation*, provide three practical basis functions to yield synthetically generated capacity fade curves. Moreover, from a top-down perspective, we argue that these parameters can also rationalize cell capacity degradation trajectories using a materials science foundation. Factors such as manufacturing variability [34] are known to affect initial capacity between cells; charging and discharging current rates may induce accelerated degradation at the end of life (EOL) or influence the linear capacity fade rate [35, 36]. Depth of discharge (DOD) and the average state of charge (SOC) during cycling may result in varying rates of capacity fade [37, 38], whereas temperature is understood to have an Arrhenius-type relationship with SEI growth-induced capacity degradation [39]. The above analysis highlights the proposed method as a top-down approach, in contrast to mechanistic modeling-based synthesis methods, which adopt a bottom-up strategy by deriving macro-level synthesis from microscopic degradation analysis at the electrode level. Unlike conventional bottom-up approaches requiring half-cell OCP data, the proposed data synthesis method generates data directly from capacity degradation curves, making it more accessible for users in field operations.

In our methodology, the *Offset parameter* governs cell-to-cell manufacturing variability, the *Slope parameter* is related to the rate of capacity fade, and the *Elongation parameter* determines the occurrence of the 'capacity knee' and the EOL. In contrast to describing a capacity degradation curve in a polynomial or exponential functional form, these three parameters can capture several properties that are relevant to LIB cell degradation. This is a favorable quality of our proposed method, as it provides an additional level of understanding of the synthetic curve.

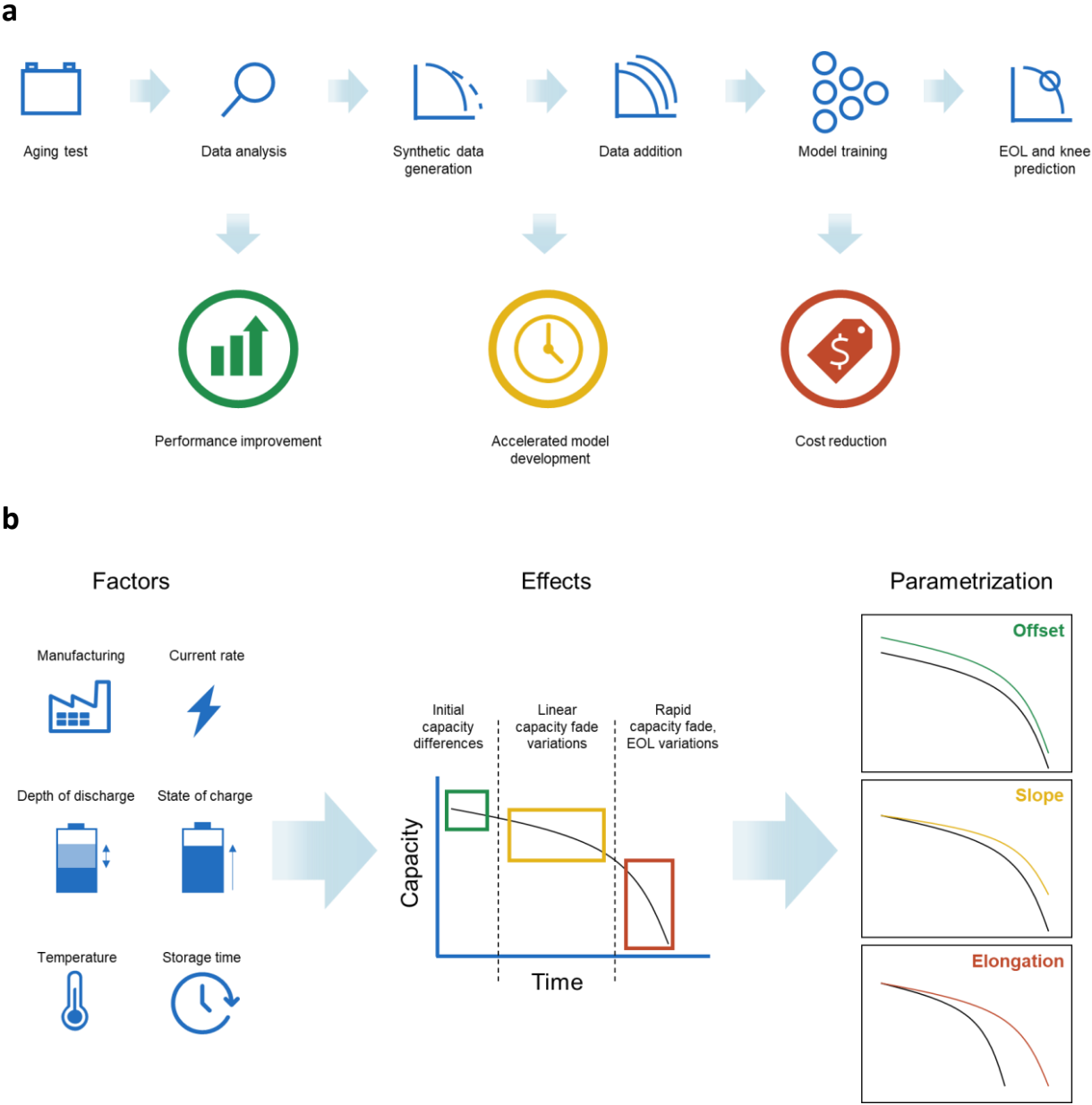

**Figure 1. a,** Framework of synthetic data augmentation based on limited battery degradation dataset. **b,** Factors influencing cell aging and the individual effects of the corresponding synthetic data generation function parameters on the modification of the seed data.

A graphical description of this method can be found in Figure 1b. First, the *Offset parameter* is related to manufacturing-induced cell variability, typical in commercial LIB cells. Commercially available LIB cells, such as the ones used in this study, are commonly assembled in a cylindrical, pouch, or prismatic form factor. These specific form factors, which have different rates of production and manufacturing control tolerances, result in characteristically different levels of cell-to-cell variations. Such variances can be manifested in discrepancies in cell capacities, internal resistances, or loading ratios. The underlying reasons for these variations could come from inhomogeneities in the electrode structure, inequitable distributions of electrolytes across the cell's volume, and varying thickness of dried electrode material - among many other causes, driven by uncertainty levels in the manufacturing process [40, 41]. Studies on large numbers of commercially available LIBs, such as 1100 LFP/Graphite cells from two different manufacturing batches, show that typical capacity and cell resistance properties can be modeled with a Gaussian distribution, with a coefficient of variation of roughly 0.5% [42].

The *Slope parameter* determines the attenuation or amplification of the capacity fade rate on a per-cycle basis. While the *Offset parameter* is mainly characterized by the manufacturing variances, the *Slope parameter* is related to degradation processes occurring due to the cycling of the cell. Throughout the life of a LIB cell, capacity degradation is often attributed to the Loss of Lithium Inventory (LLI) and the Loss of active material on positive electrode (LAM_PE) and negative electrode (LAM_NE) [43, 44]. For graphite-based LIB cells, the LLI, a hypernym for processes that lead to lithium ions becoming unable to participate in electrode insertion, acts as the most significant aging mechanism, whereas the impact of LAM_NE typically lower compared to both LLI and LAM_PE [45]. Different causes of LLI include the formation of solid electrolyte interphase (SEI), lithium becoming electrically isolated when the lithiated active material loses electrical contact and the event of irreversible plating on the graphite. Besides LLI, the rise of impedance will affect the cell's capacity, especially when the capacity is assessed with a discharge cut-off condition. Cell impedance can rise due to the thickening of an SEI layer on graphite particles and the formation of resistive layers on electrode particles [46, 47]. Both LLI and impedance increase are not strictly modeled as having a linear relationship with cycling. Yet, several studies have characterized the LLI mechanism with near-linear fade trajectories [36, 48]. LAM_PE happens due to structural degradation, particle cracking, and dissolution of active materials in the positive electrode. The LAM reduces the electrode's ability to store and release lithium ions. When LAM and LLI happen together, this typically causes an accelerated aging trend, which indicates an amplified *Slope parameter* [49]. The method proposed here amplifies or attenuates the slope of capacity fade with a multiplication factor without implying an assumption about the linearity of the underlying capacity fade curve.

Lastly, the position of the capacity knee is primarily influenced by the *Elongation parameter*, as it describes the number of cycles until the cell reaches its EOL. The capacity knee refers to accelerated degradation that is observed in LIBs [50, 51]. It has been reported that the acceleration of capacity fade can be due to the increased likelihood of plating lithium on the graphite anode as kinetic properties deteriorate at the anode. Furthermore, the nonlinear decrease of capacity can become apparent as resistance growth limits the voltage-based operating window to the flatter region of the OCV curve [52]. Regardless of the exact mechanism, the *Elongation parameter* can set the maximum lifetime of a cell, therefore implicitly determining the accelerated rate of capacity degradation past the knee point.

As we just explained, the three parameters modify the reference capacity fade curves from the seed data to create the synthetic data curves. The range from which these parameters are sampled is based on the observed statistical characteristics of the seed dataset and is further informed by empirical knowledge of battery aging mechanisms. By tailoring the variations within physically meaningful ranges, the generated synthetic data retains realistic and physically plausible degradation patterns, ensuring that the synthetic curves accurately represent the types of variations seen in real-world battery performance. This domain-specific approach ensures that the synthetic data reflects the inherent variability in battery degradation and avoids the risk of creating unrealistic or arbitrary data. The resulting synthetic curves were combined with the real capacity degradation curves as training data for the prediction models. An example of the effect on the nature of the generated synthetic data through quantitative variation of the three parameters can be seen in Figure S3 (in the supplementary materials).

Machine learning models require manual feature engineering to extract relevant features from the data. In contrast, deep learning models overcome these hurdles by extracting features from the data at the cost of large numbers of parameters [53]. To demonstrate the viability of our synthetic data generation method, we use two prediction models - a deep learning model, a convolutional neural network (CNN), and a shallow machine learning model, Gaussian process regression (GPR) - to predict the EOL and knee-point of LIBs. The input to these two models is solely the capacity of the cells, without incorporating additional features. The objective is to analyze the impact of the generated synthetic data on the models' performances and the differences observed upon drawing performance comparisons between machine- and deep-learning models for each scenario across various datasets. Five datasets that represent phenomena observed in real-world battery applications were used in the validation of the methods described in this paper. These datasets differ in cell composition, size, test conditions and aging behavior. Their descriptions have been summarized in Table 1. Additional details related to the used datasets, data pre-processing, selection of sizes of the training subsets, prediction models, model architecture and training have been elaborated in Methods.

| Dataset | Year | Number of cells/packs | Chemistry | Aging condition | Dataset characteristics |
|---|---|---|---|---|---|
| RWTH [19] | 2014 | 48 | NMC/graphite | Single test condition | Intrinsic cell-to-cell variances due to manufacturing processes |
| Stanford [54] | 2020 | 45 | LFP/graphite | Multiple fast-charging protocols | Degradation variances due to variances in stress factors |
| Oxford [55] | 2017 | 8 | LCO-NCO/graphite | Single test condition | Sparse dataset |
| NASA [56] | 2007 | 4 | - | Single test condition | Capacity recovery |
| Field [57] | 2023 | 20 | LFP | Electric vehicle | Real-world field dataset |

**Table 1.** Datasets used for method validation.

## Results and discussion

The results of this paper provide insights into the effect of varying proportions of synthetic data on the performance of the prediction models. Various training datasets were constructed through two approaches: (a) augmenting seed data with synthetic data and (b) partially replacing seed data with synthetic data. Consequently, the training datasets contained varying proportions of synthetic data. It is important to note that model validation was performed exclusively using real cell data, while synthetic data were utilized solely to enhance the training process. Initially, model performance was evaluated by training on the larger datasets, specifically RWTH and Stanford, to assess the data requirements of both shallow machine learning and deep learning models. This evaluation involved increasing (a) the number of cells and (b) the amount of cycle data available for predicting EOLs and keen points. To validate the synthetic data generation method, training subsets with varying numbers of real cells were randomly selected from the seed data, with and without synthetic data augmentation. Separate prediction models were trained on each subset to examine the consistency of predictive performance between synthetic and seed data. This analysis aimed to determine whether synthetic data could effectively substitute for seed data. Finally, the applicability of the synthetic data generation method was further explored by augmenting smaller, sparser datasets, including the Oxford and NASA datasets, as well as real-world field data. These experiments demonstrated the method's capacity to enrich limited laboratory datasets and improve the predictive performance of models trained on real-world data by incorporating high-quality synthetic data.

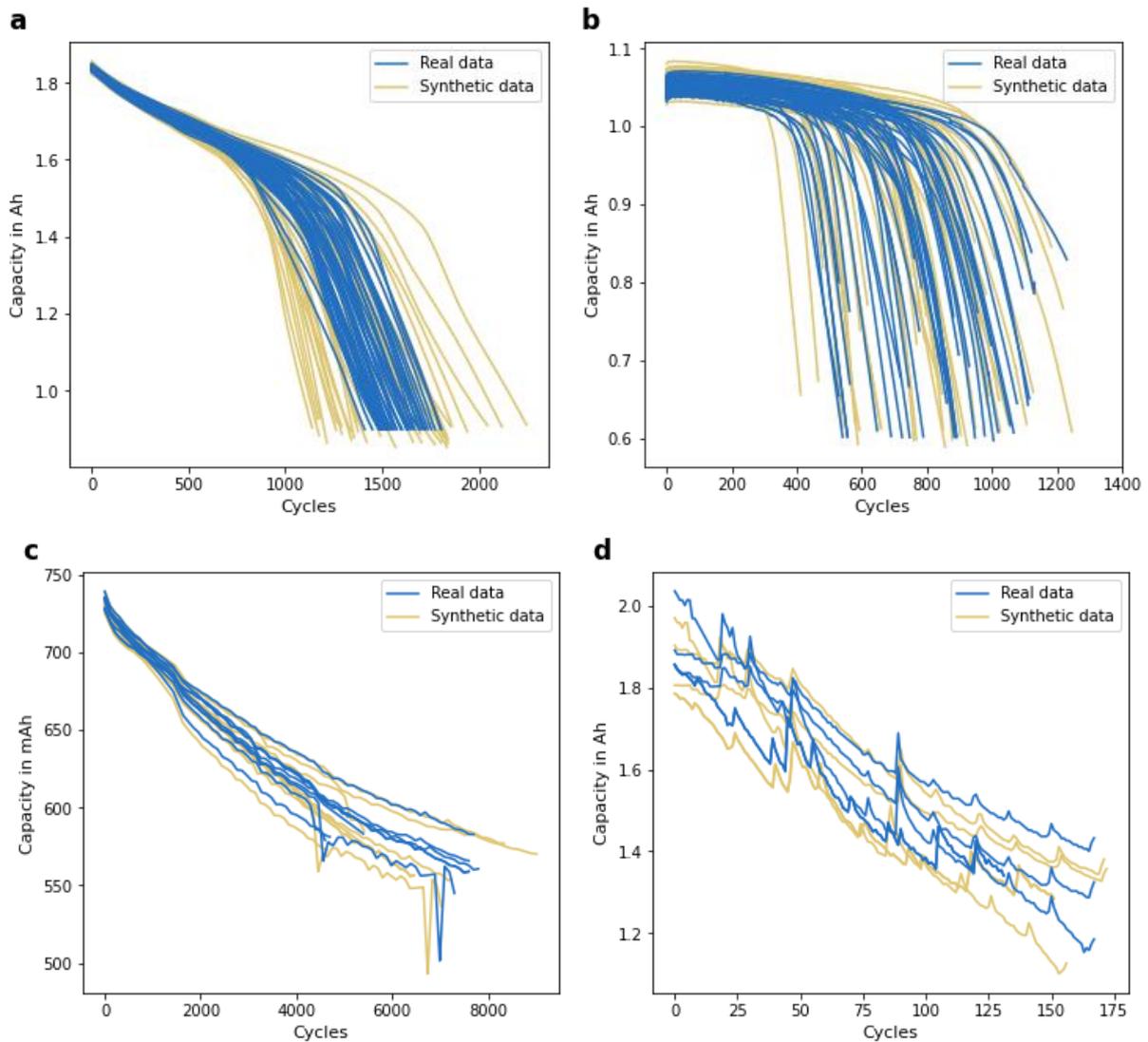

**Figure 2. a,** RWTH, **b,** Stanford, **c,** Oxford and **d,** NASA datasets along with their sample synthetic curves.

**Data analysis**

Selecting the range of synthetic data generation parameters is crucial to obtaining good-quality synthetic data. The range of these parameters should reflect the variation in stress factors, e.g., DOD, average SOC, current rate, and temperature, unique to the dataset. The cell-capacity degradation curves of the used datasets in this work and corresponding sample synthetic curves generated by our method are shown in Figure 2. Notably, the Stanford dataset exhibits an initial capacity increase, likely attributable to battery anode overhang effects. Additionally, capacity regeneration events are observed during cycling in the Oxford and NASA datasets. The synthetic curves generated by our method successfully replicate these subtle fluctuations, demonstrating the method's ability to adjust the three key parameters that characterize the shape of the capacity fade curve. In our study, an initial analysis was performed on the larger RWTH and Stanford datasets to study the distribution of the three synthetic data generation function parameters to define suitable ranges of the parameters. The histograms obtained in Figure S5 (in the supplementary materials) are the results of the pairwise calculation of the parameters. They are a normal distribution for the RWTH dataset and a flatter, more uniform distribution for the Stanford dataset, which is significantly harder to characterize. The RWTH dataset has been subjected to a single test condition and consists of data concentrated within a narrow region, while the Stanford dataset results from different conditions, resulting in a more extensive spread of the degradation data. This difference in the distributions, combined with our intention to produce significant numbers of synthetic data beyond the regions of the seed data, led to the selection of the uniform distribution for random selection of synthetic data generation function parameters during the data generation step. Additional information on the correlations between the EOL and knee-points of the real and synthetic data for the RWTH and Stanford datasets can be visualized in Figure S4. The Pearson correlation coefficient for the EOL and knee-points of both real and synthetic data of the RWTH dataset is the same ($r$ = 0.96), while that for the Stanford dataset is $r$ = 0.99. This also shows a strong similarity in the characteristics of the seed and synthetic data.

**Model training with seed data**

The analysis of the performances of the CNN and GPR models has been divided into two groups. The first focuses on adding a fixed number of synthetic data to varying numbers of seed data. In contrast, the second group shows the effect of adding synthetic data, whose numbers are multiples of the number of real capacity fade data used for model training in the respective experiment cases. The model performances in all experiments in this paper have been evaluated by calculating the averaged mean prediction error of models on all test cells in each simulation run. This paper also treats a decrease in the error as equivalent to an increase in model performance. Further details on this evaluation have been discussed in Methods. Only the results of the RWTH and Stanford datasets are presented in this section.

Figure 3 shows the performance of the CNN and GPR models using seed data purely. The results highlight how increasing the availability of real cycle data significantly impacts the prediction accuracy of both models, particularly the CNN model. The EOL prediction error band of the CNN model varies from 13.6% at input availability of 100 cycles to 5% at inputs available till 800 cycles for the RWTH dataset, as shown in Figure 3a. Figure 3e shows the respective error band from 12.8% to 7.8% for the GPR model. Furthermore, the increase in the number of real cells from 3 to 30 leads to a decrease in the average model prediction error on all test cells from 13.6% to 12.9% and from 9.4% to 5% at 100 and 800 cycles of input availability, respectively for the CNN model. For the GPR model, the corresponding errors decrease from 12.8% to 12.2% at 100 cycles and from 11.7% to 7.8% at 800 cycles. Similar trends can be observed in the Stanford dataset. As shown in Figures 3c and 3g, the band of EOL prediction errors on the test cells varies from 24.7% (100 cycles) and 8.4% (400 cycles) and from 19.8% (250 cycles) to 17.5% (400 cycles) for the CNN and GPR models, respectively. The results for the knee-point prediction errors can be analyzed in a similar fashion.

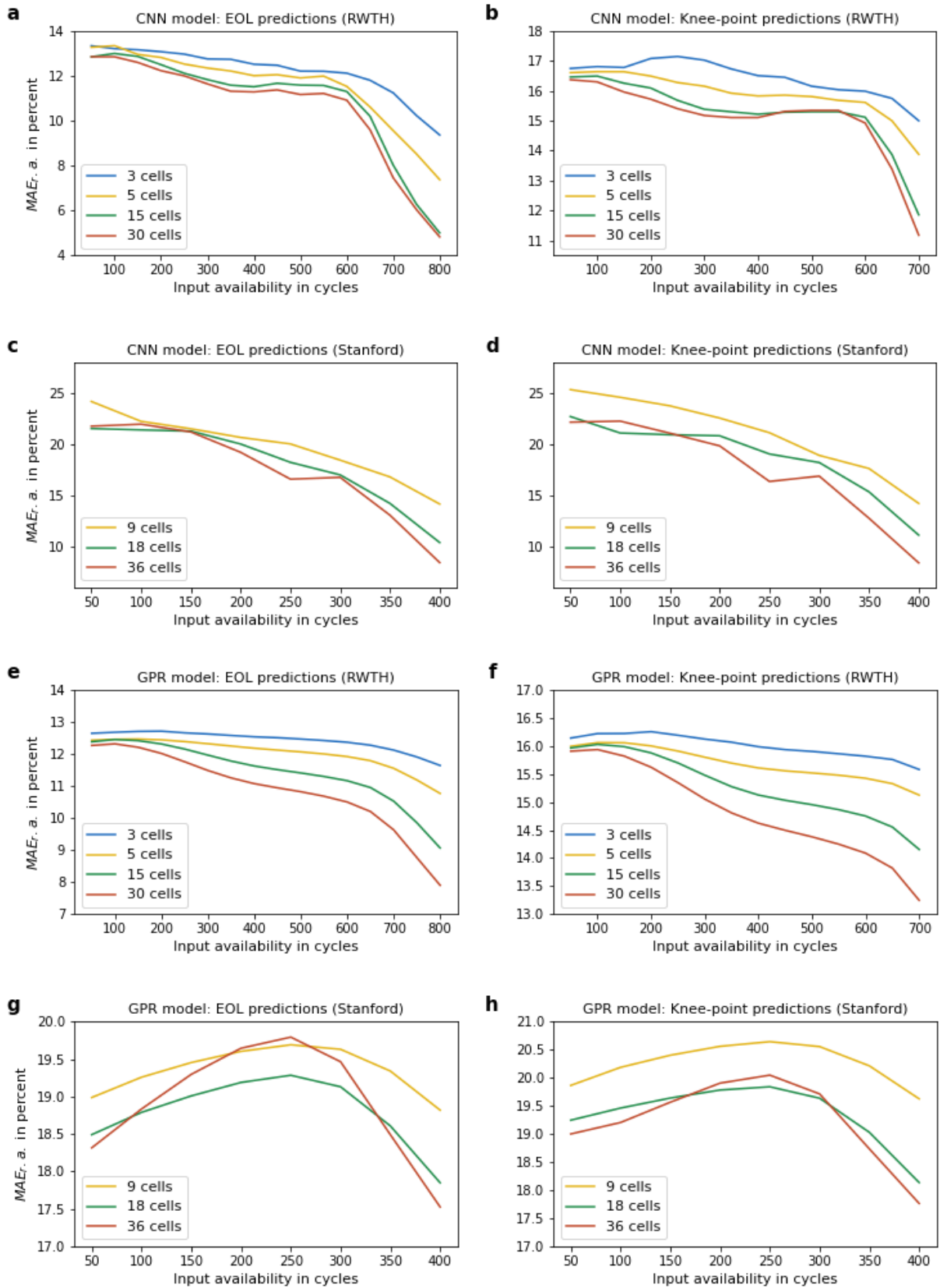

**Figure 3.** Performance of the CNN and GPR models using training data with increasing numbers of only real cells of the RWTH and the Stanford datasets. As can be seen, the performance of the CNN model increases with an increase in the number of real cell data, thereby validating the expected model behavior. The GPR model performs better with an increase in the number of cells of the seed data. However, it cannot track the EOL or the knee-point as well as the CNN model with increasing input availability in terms of cycle data and the number of cells from the seed datasets. The predictions in these plots are shown from cycle number 50.

It can be observed that the CNN model performs better with an increase in the amount of data in terms of the number of cells and the available cycle data. However, for the GPR model, performance worsens with increasing data in terms of the numbers of cells and increasing cyclic input availability in some cases, as is evident from Figures 3g and 3h in the Stanford dataset. There can be several reasons for these results. Battery capacity degradation data is highly nonlinear, and the relationship between the initial cycles up to the EOL or knee-point is complex and cannot be defined by a specific or a combination of mathematical functions. Since the GPR is predominantly function-based, it is possible that it cannot map the distribution of functions over the nonlinear data and model it as accurately as neural networks. The CNN comprises tens of thousands of trainable parameters, which are finely tuned during model training. The resulting model is better equipped to handle the complex nature of battery degradation. Another reason can be observed from the nature of the two datasets. It can be clearly observed in Figure 2 that there are no correlations between the initial capacities, the degradation trajectories and the EOL and knee-points of the cells. The RWTH dataset has far less spread in the data under a single aging profile than the Stanford dataset. The number of cells per aging profile in the RWTH dataset is also much higher (48 cells) than in the Stanford dataset (9 cells). This difference in data is also a factor responsible for irregular prediction patterns for the Stanford dataset. It may be noted that measures were taken to prevent overfitting in the CNN model, and hence, it can be ruled out during the discussion of the results. The model training has been discussed in detail in Methods.

**Addition of synthetic data**

Figure 4 shows the results from the first group of experiments by adding synthetic data in model training. With an increase in synthetic data, both models show improvements in their performance on both datasets. The largest performance increment can be observed in the cases with the least number of real cells, which can be derived from Figures 4a and 4d. The EOL prediction errors on the test cells of the RWTH dataset are similar at input availability of 100 cycles but reduced to 4% and 8.5% for the CNN and GPR models, respectively, at 800 cycles after the addition of synthetic data. For the Stanford dataset, as shown in Figure 4d regarding EOL, the errors at 100 cycles are 24.7% and 18.6% for the CNN and GPR models, respectively, which drop to 20.4% and 18% after adding synthetic data. The errors at 400 cycles reduce from 14% to 9.1% after adding synthetic data for the CNN model, while the reduction for the GPR model is observed from 18.5% to 16.2%. Taking into account the remaining figures, the maximum difference in prediction errors for the RWTH dataset at cycle 100 is negligible for all models, while those at cycle 800 are 1% each in Figure 4b (EOL) for both the models before and after synthetic data addition in the model training sets. In the case of the Stanford dataset, the maximum difference in prediction errors at cycle 100 is 3% for the CNN and 0.1% for the GPR model, as shown in Figures 4e and 4f, while those at cycle 400 are 1.8% and 1% in Figure 4e for the respective models.

The improvement in performance for most cases after adding synthetic data is much smaller for the GPR model than for the CNN model. This is because battery data is complex and nonlinear, as is the correlation between the initial capacity data and knee points or EOL. Machine learning models are simpler than deep learning models. The latter can extract low and high-level features from the complex battery data by virtue of their layered structures. An optimal number of layers, as well as the hidden neurons per layer, makes the deep-learning CNN model robust to moderate changes within the input training data. Additionally, for all cases, the CNN model, on average, performs better than the GPR model if the numbers of real and synthetic data are sufficiently large. This is because the performance of deep learning models increases with increased training data.

While the errors displayed in the results of this paper show only slight improvements in model performance, a few points need to be considered. First, these are the averaged errors of at least 15 simulation runs, and each simulation's error is an average of the prediction errors (MAE) of all test cells in the respective dataset. This double average marginalizes the errors to provide a broader view of the model performance rather than model performance on individual test cells or a single simulation. Second, the test cells for each dataset were chosen randomly without deliberately picking the cells that would showcase the best performances of prediction models. This is evident from the errors for individual test cells in Figures S9 - S12, in which some cells show extreme deviations in error, from as low as 6% to as high as 60% at an input availability of 100 cycles across all the figures. It was further verified through cross-validation, as shown in Figure S16, where different sets of test cells were selected randomly, upon which the model prediction errors were recorded by training on the remaining cells in the respective seed datasets.

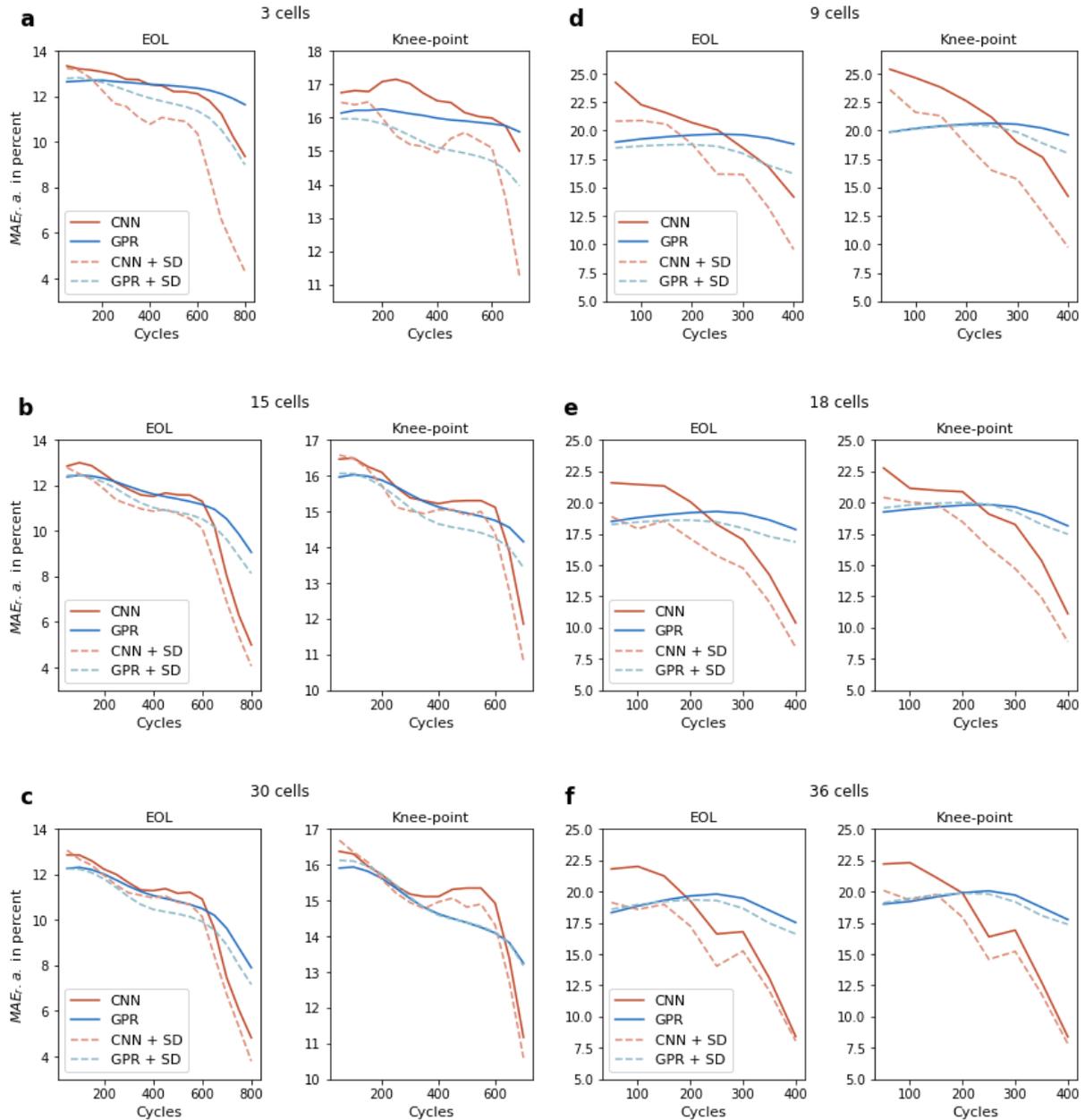

**Figure 4.** Model performances upon the addition of a fixed number of synthetic data. 30 and 36 synthetic cells were added to the model training sets for the RWTH (**a**, **b**, **c**) and the Stanford (**d**, **e**, **f**) datasets. 'SD' in the legend stands for 'Synthetic data', indicating the presence of synthetic data in the training sets. In every subsequent row of graphs, there is an increase in the number of real training cells, which is the same as the number of cells used as the base for synthetic data generation for that case. Performance comparisons of both the GPR and the CNN models have also been highlighted.

**Partial replacement of real data with synthetic data**

The ability of synthetic data to replace real data while not impacting model performance negatively is perhaps one of the clearest validations of any synthetic data generation method. In this study, we systematically evaluated how much real data can be replaced by synthetic data while maintaining model performance. Figure 5 shows the respective performances of the CNN and GPR models for the EOL predictions. The results demonstrate that for both the RWTH and Stanford datasets, up to 50% of the real training data can be replaced with synthetic data without negatively affecting the CNN model's prediction accuracy. Specifically, the model's performance with a combination of synthetic and seed data is either better than or as good as with purely real data for the CNN model. Better model performance can be attributed to the random synthetic data generation within the tolerances defined by the synthetic data generation function parameters, which yielded some data that had similar features as the unseen test data. The same can be seen for the CNN model's performance on the knee-point predictions for the test cells in Figure S6. However, for both the EOL and knee-point predictions in Figure 5 and Figure S7, respectively, the performance of the GPR is inconsistent. The GPR model can track the EOL using training sets containing synthetic data to a limited extent as opposed to training sets containing purely real data. This can be observed in Figures 4e, 4f, 4g, and 4h, where the green line graphs, corresponding to mixed training sets, closely follow the red line graphs, corresponding to purely real training sets, only in some areas. In the case of knee-point predictions, as shown in Figure S7, the model performances with the mixed training sets are even worse. However, these results are similar to those obtained using only real data, as shown in Figures 3g and 3h. This suggests that the CNN model can generalize well even when trained with synthetic data, while the GPR model struggles due to its inherent limitations rather than the quality of the synthetic data.

The discussed method was able to generate degradation data that is different from the already present training data without the expenditure of time and resources for cell testing, which is desirable and the intended purpose. Results from Figure 5 also serve as a base for calculating the effort reduction regarding the number of cells being tested, which is proportional to the shares of synthetic data within the respective training sets. Based on our studies, it is possible to reduce cell-testing effort by at least 50% using the synthetic data generation method for similar performance to our prediction models. These results highlight the practical advantage of synthetic data generation, as it can significantly reduce experimental workload without sacrificing model accuracy. For an example of testing effort reduction calculations, the readers are directed to Note S1 in the supplementary material.

The readers are reminded that for all the experiments performed in this paper, the selection of training subsets of seed data and the generation of synthetic data from those subsets were completely random. Therefore, with more precise selection of seed data and stricter control over the synthetic data generation process, further improvements in model performance are expected.

There have been some common observations regarding the prediction results of the two datasets used so far. The prediction errors of EOL and knee-point in the Stanford dataset were significantly larger than those in the RWTH dataset. This is because the Stanford dataset demonstrates a much larger variation in terms of the EOL and knee-points of cells due to different internal processes taking place within the cells under stress factors, e.g., current rates. The nine aging protocols themselves are significantly different, resulting in different types and extents of degradation of the cells. Prediction errors were also observed to be higher for knee-point than EOL. Knee points are not precisely defined as sharp points on capacity fade curves. Hence, differences can exist in the ways that the prediction models interpret them during training compared to the well-defined EOL.

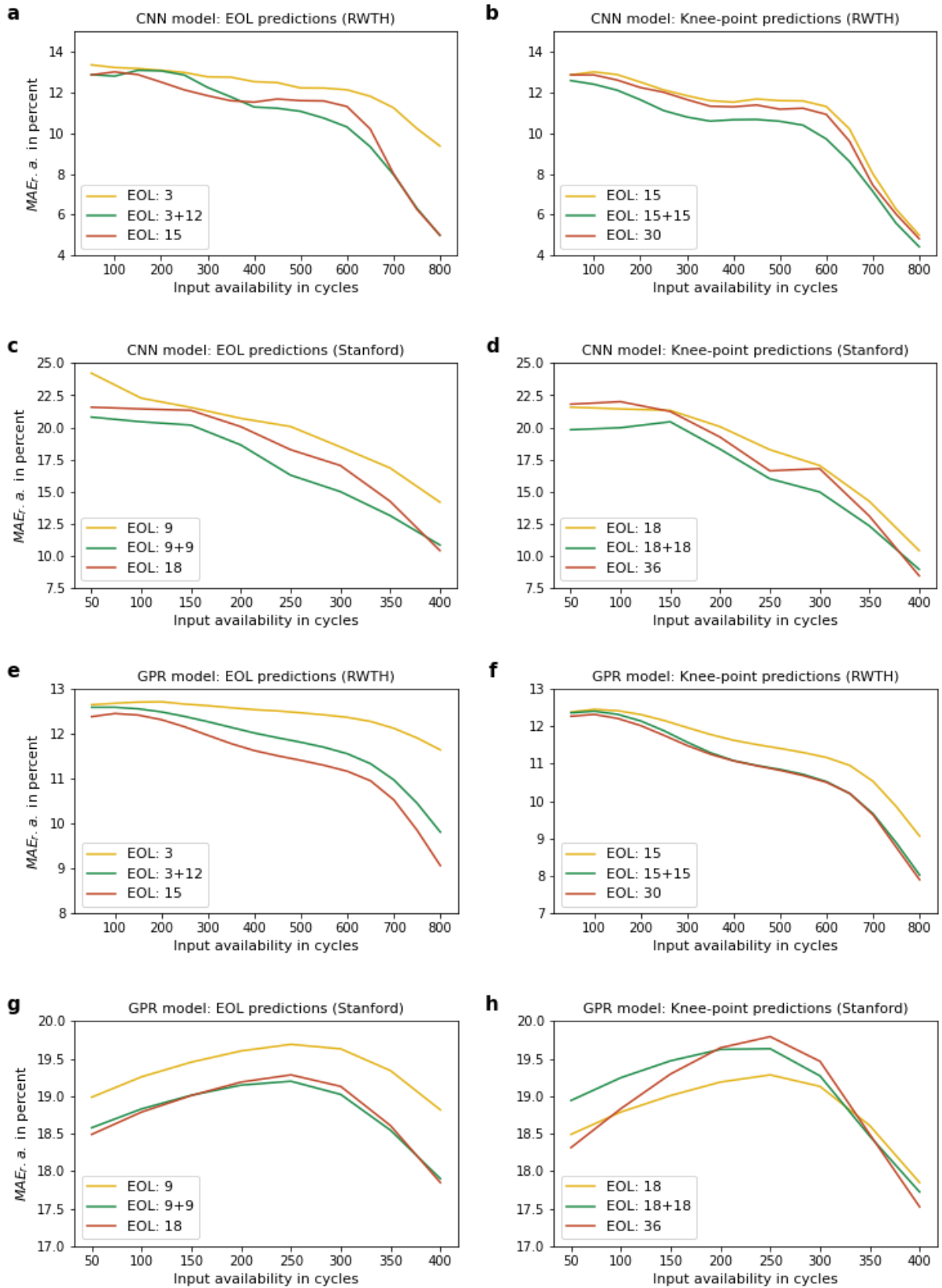

**Figure 5.** Replacement of real data by synthetic data for EOL and knee-point predictions by the CNN and GPR models for the RWTH and the Stanford datasets. Error line plots for model performance on training sets having only real curves are represented in the legend as a single integer 'n', while for training sets having a combination of 'n' real and 'm', synthetic curves are represented as 'n+m'. The predictions in these plots are shown from cycle number 50.

The advantages of our synthetic data generation method are observed predominantly due to the defined ranges of the synthetic data generation function parameters: Offset, Slope, and Elongation. The selection of the parameters within these ranges was performed randomly to avoid the influence of human bias on the type of data generated and, hence, the performance of the prediction models. Further details on the statistical and sensitivity analysis can be found in Methods. Having said that, the synthetic data generation function can be used to produce the exact nature of synthetic degradation curves required for different applications as well as to improve model accuracy significantly, and the statistical analysis step is not mandatory. However, these analyses help address the concerns regarding the extrapolation capability of our synthetic data generation method beyond the boundaries of the seed dataset and the real-world significance of the generated data. A significantly large extrapolation of synthetic data beyond the boundaries of the seed dataset may not be a valid real degradation path for cells under the specified aging conditions. As visible from the seed datasets and their respective sample synthetic curves in Figure 2, the statistical analysis step aids in selecting parameter ranges that result in creating the synthetic curves within and slightly beyond the bounds of the seed datasets.

**Validation on sparse datasets**

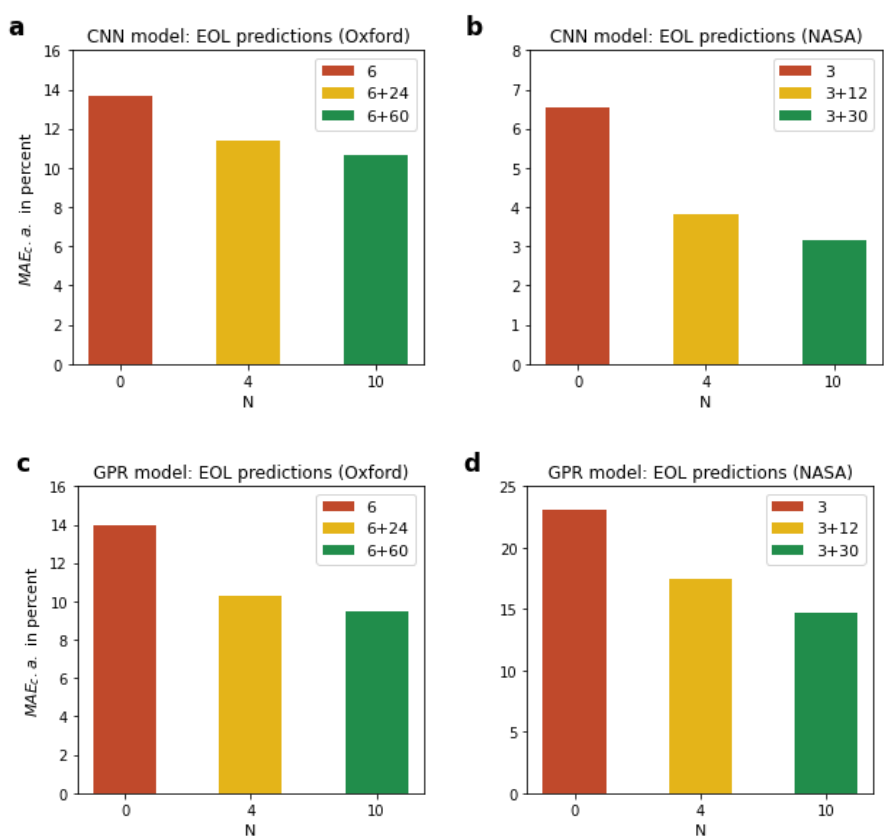

**Figure 6.** Mean errors, *MAE*$_{c.a.}$ for performance validations of the CNN **(a, b)** and GPR **(c, d)** models on the Oxford and NASA datasets. Error bars for model performance on training sets with only real curves are represented in the legend as a single integer 'n', while for training sets with a combination of 'n' real and 'm', synthetic curves are represented as 'n+m'. 'N' denotes the ratio of the number of synthetic curves to the number of real curves within the model training set for the respective cases. Figure S15 (in supplementary materials) shows the evolution of errors with cycle numbers in the form of line graphs. It may be argued that the GPR model performs better than the CNN model for the Oxford dataset. Some plausible reasons could be the nature of the dataset in terms of complexity, the number of cells, and the training mechanism of the two models on the dataset. However, the primary objective of the paper is to showcase the effect of adding synthetic data on the prediction accuracy of the machine learning and deep learning models and not draw comparisons between the two models themselves. A detailed comparative analysis of the performances of the two models is, hence, out of the scope of this paper.

The common range of the elongation parameter obtained from the sensitivity analysis was used for the performance validation of the two models on the sparse Oxford and NASA datasets. Figure 6 shows the mean errors of model performances upon training sets having varying numbers of synthetic data added to seed data. The figures depict the errors *MAE<sub>r.a.</sub>* and *MAE<sub>c.a.</sub>* as described in equations (16) and (17) in Methods. With the increase in the proportion of synthetic data within the training subset, the performances of both the CNN and GPR models also increase. The Oxford dataset has limited cells tested under a single condition but displays much larger cell-to-cell variations than the RWTH dataset. The capacity degradation curves of the NASA dataset show capacity recovery at several stages of degradation. Therefore, it is evident that the proposed synthetic data generation methodology helps generate data resembling actual cell data that can enrich smaller datasets while improving the predictive performances of both machine- and deep-learning models.

**Validation on field datasets**

To validate the performance of the proposed method using field data, we analyzed degradation data from battery packs in 20 commercial electric vehicles operating in real-world conditions over two years [57]. Field data presented significant noise due to measurement uncertainties, a challenge not present in laboratory datasets. Capacity recovery observed during long rest periods between driving profiles posed an additional challenge for data augmentation. Figure S17 displays the original field data alongside synthetic data generated by the proposed method, distinguished by different colors. To validate the proposed data augmentation approach on field datasets, we randomly selected data from 15 vehicles as seed data to generate synthetic data. The remaining data from the other five vehicles were used for validation. Figure 8 shows the mean errors of model performances with varying amounts of synthetic data added to the seed data in the training sets. As the proportion of synthetic data increased, the performance of both the CNN and GPR models improved. The models' mean errors decreased upon integrating synthetic data into the training set, thereby verifying the effectiveness of the proposed synthetic method for enhancing lifetime prediction in complex, noisy battery field data.

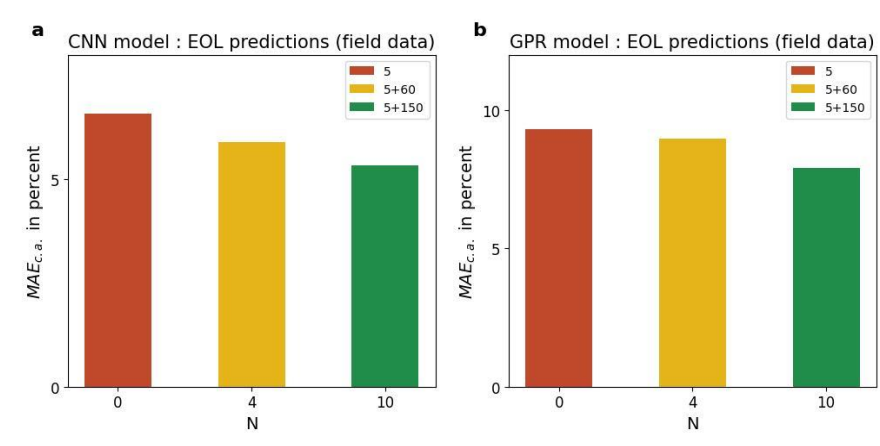

**Figure 8.** Mean errors, *MAE<sub>c.a.</sub>* for performance validations of the CNN **(a)** and GPR **(b)** models on the field datasets. Error bars for model performance on training sets with only real curves are represented in the legend as a single integer 'n', while for training sets with a combination of 'n' real and 'm', synthetic curves are represented as 'n+m'. 'N' denotes the ratio of the number of synthetic curves to the number of real curves within the model training set for the respective cases.

**Applications and outlook**

The primary applications of the work of this paper are lifetime prediction of LIB systems, reduction of resources for cell testing, and enrichment of operation datasets when having insufficient data. The proposed synthetic data generation process can help increase the quantity and quality of the sparsely available field operation data while decreasing cell testing effort by at least 50 percent. Additionally, this can lead to better training of prediction models using field data and their integration into battery management systems. Even though the work in this paper randomly selected the cells for each seed dataset, it could be possible to train models for specialized applications by selecting specific capacity degradation curves in the seed dataset. Moreover, the described methodology of synthetic data generation and deep-learning neural networks can be adapted and applied to data-driven studies in entirely unrelated domains of research that share similarities with the non-linearity observed in LIB systems.

However, several challenges continue to exist for the accurate prediction of LIB degradation. To minimize the number of cell aging tests and to generate relevant synthetic data catering to a wide range of field applications, the aging protocols and the number thereof must be well-defined. It is known that capacity degradation in LIBs is correlated to an increase in their internal resistance. Thus, including resistance data as an additional feature to prognostic models could increase degradation prediction accuracy. Hence, generating synthetic resistance curves correlated to the corresponding synthetic capacity degradation curves for model training would be extremely advantageous. Finally, even though the proposed model has been trained on different datasets with characteristics that resemble many aspects of field data, and the results obtained show promise, the actual implementation of the synthetic data generation method and subsequent model training and validation on field data is still needed.

## Conclusions

In this paper, we developed a method of fast data augmentation with the objective of increasing the predictive performance of data-driven models while reducing cell testing efforts and costs. Three parameters are designed based on the degradation mechanisms to introduce the variations in degradation trajectories without complicated, time-consuming electrochemical simulations. We demonstrated its advantages in conjunction with both machine learning and deep learning models to predict the occurrences of the end-of-life and knee-points of battery degradation curves. Four datasets from different chemistries, which exhibit vast differences in terms of cell capacities, inter-cellular variations, testing protocols, and the presence of measurement noise and capacity recovery, were used to validate our method. The addition of synthetic data showed an increase in prediction accuracy, while the partial replacement of real data with synthetic data also led to similar model performances using limited cell data. Performance comparisons were also drawn between shallow machine learning and deep learning models. From a broader perspective, the work of this paper opens possibilities to extend its application to battery health prognostics using sparsely available field data from energy storage systems that have larger variations and more relevance to practical applications than laboratory-generated test data.

## Methods

### Synthetic data generation function

The offset parameter is described as the difference between the initial capacity values of two cells in Ah. The slope parameter, represented in Ah, is determined by the variances in the slopes of the capacity fade curves within the initial linear degradation phase, in which the spread in capacity degradation is relatively less yet prominent. It is calculated by first subtracting the initial capacity values of the two cells, followed by subtraction of the differences in capacities at the $n$th cycle of the cells up to which minor differences in slopes are observed. The elongation parameter is defined as the ratio of the cycle numbers of the endpoints of two LIB degradation curves and is a dimensionless quantity. If $o$, $s$ and $e$ represent the offset, slope, and elongation parameters, respectively, then

$$o = Q_{i,1} - Q_{j,1} \quad (1)$$
$$s = (Q_{i,n} - Q_{i,1}) - (Q_{j,n} - Q_{j,1}) \quad (2)$$
$$e = N / M \quad (3)$$

where $i$ and $j$ represent any two cells taken at a time for the calculation of the parameters that have initial capacities $Q_{i,1}$ and $Q_{j,1}$, capacities at the $n$th cycle $Q_{i,n}$ and $Q_{j,n}$ and cycle numbers of their final data points $N$ and $M$, respectively.

A cell's capacity data can be represented in the form of a capacity vector $Q$ and the corresponding cycle number vector $C$, both being of the same length $L$. The vectors of the corresponding parameters, having lengths equal to $L$ are then prepared as follows.

$$O = (o, \dots, o) \quad (4)$$
$$S = (0, \dots, s) \quad (5)$$
$$E = (1, \dots, e) \quad (6)$$

where $O$ is the offset parameter vector and is a constant vector. $S$ and $E$ are the slope and elongation parameter vectors that are obtained by linear interpolation between the initial and final values, as shown in equations (5) and (6). The real curve is combined elementwise with these parameter vectors to obtain the modified data as

$$Q' = Q + O + S \quad (7)$$
$$C' = C * E \quad (8)$$

The modified cycle-number vector $C'$ will comprise non-integer values. Hence, a new vector $C_s'$ is generated by linear interpolation of the integer values between the initial value of $C$ and the final value of $C'$ rounded off to the closest integer. $C_s'$ is the cycle number vector of the new synthetic curve. The corresponding synthetic capacity values $Q_s'$ are obtained with the interpolation of capacity values at cycle numbers $C_s'$ against the vector $Q'$. The resulting synthetic curve vectors will have lengths of $(L * E)$ rounded off to the nearest integer. The variations of synthetic data with the change in the parameters can be quantitatively visualized in Figure S3.

**Datasets**

High-quality datasets are essential for developing accurate data-driven models for battery applications. The generation of such datasets requires high-precision measurement equipment, controlled testing environments, and standardized testing protocols to minimize noise and measurement errors, thereby ensuring robust model training and validation.

In battery capacity estimation and prediction tasks, high-quality laboratory data typically involve capacity and resistance measurements obtained using high-precision sensors under consistent, fixed check-up testing procedures conducted at regular intervals. This ensures the availability of sufficient and reliable data for model development. In contrast, high-quality field data are characterized by high sampling rates, low sensor noise, and minimal logging errors, such as data gaps. In optimal scenarios, reliable capacity tests are conducted in workshops to provide accurate data labels for aging diagnosis and prognostics tasks.

The dataset generated by RWTH consists of 48 Sanyo/Panasonic UR18650E graphite/NMC cylindrical cells with a nominal capacity of 1.85 Ah aged under the same load profile and test conditions. Initial performance of the cells was ascertained with begin-of-life (BOL) tests, with regular reference parameter tests (RPT) carried out to determine cell performance. The validation batch of cells tested by Stanford consists of 45 A123 Systems' APR18650M1A graphite/LFP cells having a nominal capacity of 1.1 Ah. These cells underwent nine different fast-charging protocols, five cells per protocol, and were cycled to failure. The Oxford dataset consisted of 8 Kokam pouch cells having lithium cobalt oxide (LCO)/lithium nickel cobalt oxide (NCO) positive electrode and graphite negative electrode. The cells had a nominal capacity of 740 mAh and were aged till EOL of 80% nominal capacity under a constant current of 2 C of a CC-CV profile and discharged under an urban drive cycle profile. Characterization tests were carried out every 100 cycles at 1 C current. The NASA dataset's test batch was used, containing four cells labeled 5, 6, 7 and 18. These cells had a nominal capacity of 2 Ah and were aged under three different profiles at room temperature. Charging was performed at a constant current of 1.5 A till the cell voltages reached 4.2 V, followed by a constant voltage stage until the charging current dropped to 20 mA. The discharge was performed at a constant current of 2 A till the voltages of cells 5, 6, 7, and 18 reached 2.7 V, 2.5 V, 2.2 V and 2.5 V, respectively. The experiments were conducted until the defined EOL of 30 percent capacity degradation in all cells, i.e., from 2 Ah to 1.4 Ah. The field dataset consists of data from 20 commercial electric vehicles operating in real-world conditions over two years. Each vehicle used identical battery systems. Raw capacity data for each vehicle were obtained using a variant of the Ampere integral equation and statistical calibration to mitigate errors caused by imprecise SOC measurements and data noise. Due to the embedded BMS limitations in computation and data transmission, capacity calculations were performed monthly.

**Data pre-processing**

Pre-processing was performed for both the model train and test sets. To reduce the model training time, the input capacity curves were reduced in resolution by sampling only one in every two points. This sampling rate was chosen after analyzing trade-offs between model performances with data using different sampling rates and their respective computation times. The outputs were the knee or EOL points of the respective degradation curve. This was applied directly to the Stanford, Oxford and NASA datasets, and no additional pre-processing was required. However, the RWTH dataset required interpolations of intermediate capacity data points, which was done using the Piecewise Cubic Hermetic Interpolating Polynomial (PCHIP) function available in MATLAB. For the field dataset, due to the sparsity of monthly measured raw data for model training, we initially used cubic spline interpolation to estimate daily capacity changes.

**Calculation of synthetic data generation function parameters**

The three synthetic data generation function parameters were obtained separately for each dataset. These parameters were calculated, taking two cells at a time within each dataset and each aging test condition. The calculation of the offset and elongation parameters has already been discussed as part of the methodology. The values of $n$ chosen for calculating the slope parameter were 500, 200 and 500 cycles for the RWTH, Stanford and Oxford datasets, respectively. The NASA dataset did not have any similarities in capacities and slopes in the initial cycles of the degradation curves. Hence, the slope parameter for this dataset was assigned a value of 0. The resulting sets of these parameters for the RWTH and Stanford datasets have been represented as histograms in Figure S5. This can serve as a starting point for selecting the synthetic data generation function parameters for a given dataset to obtain representative synthetic data, as was done for the Oxford and NASA datasets. However, such an analysis is not mandatory, and the parameters can be selected based on the nature of the synthetic data required.

**Model training methodology**

For the RWTH dataset, 7 test cells were randomly selected, and the training set had 40 cells. For the Stanford dataset, one cell was randomly selected for the validation batch of cells from each of the nine charging protocols. This resulted in a total of 9 tests and 36 train cells. In the Oxford dataset, two cells were randomly selected for the test set, leaving a total of 6 cells for training. The NASA dataset had three training cells and 1 test cell. All datasets were trained separately in all the simulations. The training and test data of the datasets have been plotted in Figures S1 and S2.

The model performances for the larger datasets were observed by increasing the available real cell data and the number of available cycle data of test cells to predict their EOL and knee points. We increased the number of randomly selected cells from the available real training cell data to create training subsets with or without synthetic data and trained the prediction models separately for each subset. For the RWTH dataset, we increased the seed data in the order 3, 15 and 30 cells, whereas for the Stanford dataset, the number of cells was increased by one cell per test condition, resulting in the order of 9, 18 and 36 cells. Because these training subsets were selected at random, simulations were run at least 15 times per scenario. The average errors of these runs/trials have been presented as the results. This was done to reduce the weights of the results obtained by inadvertently selecting 'good' or 'badly' biased training subsets by the random function.

In all simulations, the numbers of both real and synthetic curves in each training subset were specifically chosen such that the minimum ratio of the numbers of synthetic curves to real curves within the subset was equal to 1. This would ensure a significant contribution to model performance by the synthetic curves. The selected numbers of real and synthetic data also prevented repetitive experiments from understanding the behavior of the prediction models in response to a wide range of training data.

### Cross validation

To ensure transparency regarding the final cell selection for model training and test sets, k-fold cross-validations were performed. Some results for the same have been provided for knee-point and EOL predictions for both the GPR and CNN models on the RWTH and Stanford datasets, whose results have been plotted in Figure S16. The value of 'k' for the RWTH and Stanford datasets chosen were 6 and 5, respectively. It can be observed that the margin of prediction errors can vary enormously based on the test set and that the errors corresponding to the test cells that were randomly selected and fixed for all the results in this paper roughly lie within the shaded regions.

### Sensitivity analysis

The synthetic data generation function takes ranges of values of the three parameters as inputs, and a random function within it chooses a set of parameter values from those ranges. These are used to generate synthetic degradation curves. A sensitivity analysis was performed alongside the statistical analyses, as shown in Figure S5, to determine the optimal ranges of parameter values. This was done to support the conclusions of the parameter ranges derived in the statistical analyses of the used datasets. It also offers insights into the pre-selection of these parameter value ranges for other datasets with very few cells, which could benefit from the generation of synthetic data, but for whom carrying out statistical analysis could lead to erroneous results due to their small size.

The sensitivity analysis was carried out by varying the ranges of one parameter while keeping the other parameters fixed. Results obtained by such analysis provided some interesting insights. Firstly, as expected, the model performance results and the process of synthetic data generation were most sensitive to the elongation parameter due to its direct impact on the cycle life of the synthetically generated cell. The variation of cycle life of cells in the datasets, and hence the elongation parameter, is the largest as compared to the offset and slope parameters. Secondly, an optimum range exists for the selection of the elongation parameter. A narrow range of elongation of ±10% was observed to produce unevenly distributed synthetic data, with a higher density of synthetic data at the extreme boundaries, leading to worse results by synthetic data addition as compared to the cases of adding no synthetic data at all. A high range of ±40% and above improved the model performance but led to the creation of synthetic data that no longer resembled the original dataset. The ranges obtained by the statistical analysis in Fig. 2 produced results as good as high parameter ranges while at the same time maintaining the likeness to the real cells within the datasets. To further generalize the conclusion of the analysis, a common range of the elongation parameter of ±25% was selected and used for all experiments and all four datasets in this paper. The offset and slope parameters did not influence the results in any clear manner and were obtained directly from the respective dataset due to their dimensionality (Ah), which would be different for different datasets. A selection of extremely high values of these two parameters would be unreasonable due to the lack of resemblance of synthetic data with the seed data and was therefore not considered.

**Convolutional Neural Network**

Among the various deep learning neural networks available, we implemented a CNN-based model for the purpose of point prediction. The core of such models is the convolutional layer, which can extract and preserve the relationships between different data points using its constituent kernels. CNNs apply the mathematical operation, convolution, to the input series $x$ and its composite kernel $w$ to produce the resulting output series $y$.

$$x = [x_0, x_1, x_2, \dots, x_{m-1}] \tag{9}$$
$$w = [w_{-p}, w_{-p+1}, \dots, w_0, \dots, w_{p-1}, w_p] \tag{10}$$
$$y = [y_0, y_1, y_2, \dots, y_{m-1}] \tag{11}$$
$$y_n = \sum_{k=-p}^{p} x_{n-k} w_k \quad \forall \ n \in [0, m-1] \tag{12}$$

In the context of our work, $x$ is one of the multiple inputs within the capacity data array created during data pre-processing, whereas $y$ is the output series of the previous layers, which is received as input by the final dense layer of the CNN model. The dense layer then performs a series of matrix-vector multiplication, producing a single point as the prediction output of the model. Our CNN model architecture consisted of convolutional, max-pooling, dropout, dense and flattened layers. The 'relu' activation function was used to improve the model's ability to learn nonlinear relations. The number of epochs for all the experiments was fixed at 700 to provide justified comparisons of results obtained from experiments on different datasets as well as the variation of input data within each dataset. The kernel size was kept to a minimum, which resulted in the best possible learning of features of the data. Two dropout layers with a dropout percentage of 10 percent each were added to the model. Additionally, a callback was used in model training to store the model with the lowest validation loss. These measures were taken to prevent model overfitting on the training dataset. A detailed structure of the CNN model used for the experiments has been shown in Figure S13.

A typical lithium-ion capacity fade curve is a time series and, simultaneously, spatial data. The 1-D CNN was chosen because of its ability to extract the spatial distribution of data as well as features with the help of kernels. The initial preference to use Long-Short-Term-Memory (LSTM) based models was intuitive due to their ability to detect temporal relationships of time-series data and their abundant utilization in existing studies and literature. However, after several experiments for the purpose of knee-point and EOL prediction, CNN-based models had a significant reduction in training time by approximately 1100-1600%, with no noticeable effect on performance.

**Gaussian Process Regression**

We used the Gaussian Process Regression (GPR) based model to provide a benchmark model for performance comparison. It is a supervised machine learning algorithm that defines a distribution over functions that can be defined over the available data. A combination of k-fold cross-validation and manual hit and trial was used to fix the GPR hyperparameters. After performance comparisons with the radial basis function (RBF) kernel, the Matérn 3/2 kernel was selected, which is given by:

$$k(x_i, x_j) = \frac{1}{\gamma(v) 2^{v-1}} \left( \frac{\sqrt{2v}}{l} d(x_i, x_j) \right)^v K_v \left( \frac{\sqrt{2v}}{l} d(x_i, x_j) \right) \tag{13}$$

where $d$ is the Euclidean distance between the points $x_i$ and $x_j$, $K_\nu$ is a modified Bessel function and $\gamma$ is the gamma function. The value of $\nu$ is 1.5 for the used kernel.

Numerous scientific works have benefited from the GPR for EOL and knee-point predictions. In all the studies, however, feature engineering and hyperparameter tuning were crucial steps. For a justified comparison of our CNN-based model performance with the GPR serving as a benchmark, no feature engineering was performed and hence, the inputs were the same for both models. This was done to highlight the learning ability of our CNN-based model without the need for feature engineering as well as changes in any of its hyperparameters.

**Identification of EOL and knee point**

The reduction of the cell's discharge capacity to 80% of its nominal capacity was this study's chosen definition of end-of-life (EOL). Among the various available knee point identification methods [16, 50, 58], the one proposed in [16] was used. It involves fitting two tangent lines to the capacity fade curves that represent the two degradation stages. The cycle number of their point of intersection is defined as the knee point. This method outperformed the others on both datasets with minimum adjustment of the function's parameters, capacity data processing, and application complexity. It could also be applied across multiple datasets having different natures of capacity fade curves.

**Evaluation metrics**

The model performance was evaluated based on the difference in the predicted and actual EOL and knee point values. The prediction error of the model for a test cell is given by

$$\Delta = y_p - y_a \tag{14}$$

where $y_p$ and $y_a$ are respectively the predicted and actual values of the knee-point or EOL of the test cell. The unit can either be the percent or the number of cycles.

However, since the focus is on the performance of the prediction models for different numbers of randomly selected cells of the datasets and randomly generated synthetic data as an ensemble, the average errors for all the test cells at the respective cycle numbers were obtained instead of individual cells. For error metrics, we chose the mean absolute error (MAE), which can be represented in cycles as well as percentages. Thus, the MAE of the entire group of test cells is defined as

$$MAE = \frac{\sum_{i=1}^{n} |y_{i,p} - y_{i,a}|}{n} \tag{15}$$

where $y_{i,p}$ and $y_{i,a}$ are the predicted and actual knee-point or EOL values of the $i$th test cell, averaged over $n$ test cells, which is 7, 9, 2 and 1 for the RWTH, Stanford, Oxford, and NASA datasets, respectively. Because of randomly selected subsets of cells for training with different cells or different combinations of randomly selected cells in every subset and the random synthetic data generated, several runs were performed. The mean absolute errors of all the *m* runs, MAE run-average (*MAE$_{r.a.}$*) is calculated as in equation (16). The error-calculation procedure has been graphically represented in Figure S14.

$$MAE_{r.a.} = \frac{MAE}{m_{runs}} \tag{16}$$

The errors for the Oxford and NASA datasets, as shown in Figure 6, represent the mean of *MAE$_{r.a.}$* over its *p* points of cycle data. This is referred to as the MAE cycle average (*MAE$_{c.a.}$*) in this paper.

$$MAE_{c.a.} = \frac{MAE_{r.a.}}{p} \tag{17}$$

Another commonly used metric is the root mean squared error (RMSE), which measures the average error and the error distribution. However, it provides an ambiguous interpretation of errors. MAE, on the other hand, is a much more reliable and intuitive measure of errors, especially when comparing model performances [59], and hence, we only use MAE as the performance metric for model performance evaluations.

## Data availability

The data used for method validation in this work are from RWTH Aachen University [19], Stanford University [54], University of Oxford [55], and NASA [56]. All pre-processed data in this study will be deposited in GitLab for public accession before the final publication.

## Code availability

All codes generated in this study will be deposited in GitLab for public accession before the final publication.

## Acknowledgment

This work has received funding from the research project "SPEED" (03XP0585) funded by the German Federal Ministry of Education and Research (BMBF).

# Fast data augmentation for battery degradation prediction

Weihan Li[1,2*§], Harshvardhan Samsukha[1,2§], Bruis van Vlijmen[3,4], Lisen Yan[1,2], Samuel Greenbank[5], Simona Onori[3,6], Venkat Viswanathan[7]

[1] Center for Ageing, Reliability and Lifetime Prediction of Electrochemical and Power Electronic Systems (CARL), RWTH Aachen University, Campus-Boulevard 89, 52074 Aachen, Germany
[2] Institute for Power Electronics and Electrical Drives (ISEA), RWTH Aachen University, Campus-Boulevard 89, 52074 Aachen, Germany
[3] SLAC National Accelerator Laboratory, Menlo Park, California 94025, USA
[4] Department of Materials Science and Engineering, Stanford University, Stanford, California 94305, USA
[5] Department of Engineering Science, University of Oxford, Oxford, OX1 3PJ, UK
[6] Department of Energy Science and Engineering, Stanford University, Stanford, California 94305, USA
[7] Department of Aerospace Engineering, University of Michigan, Ann Arbor, Michigan 48109, USA

* Correspondence: weihan.li@isea.rwth-aachen.de (W.L.)
§ Both authors contributed equally

Supplementary figures:

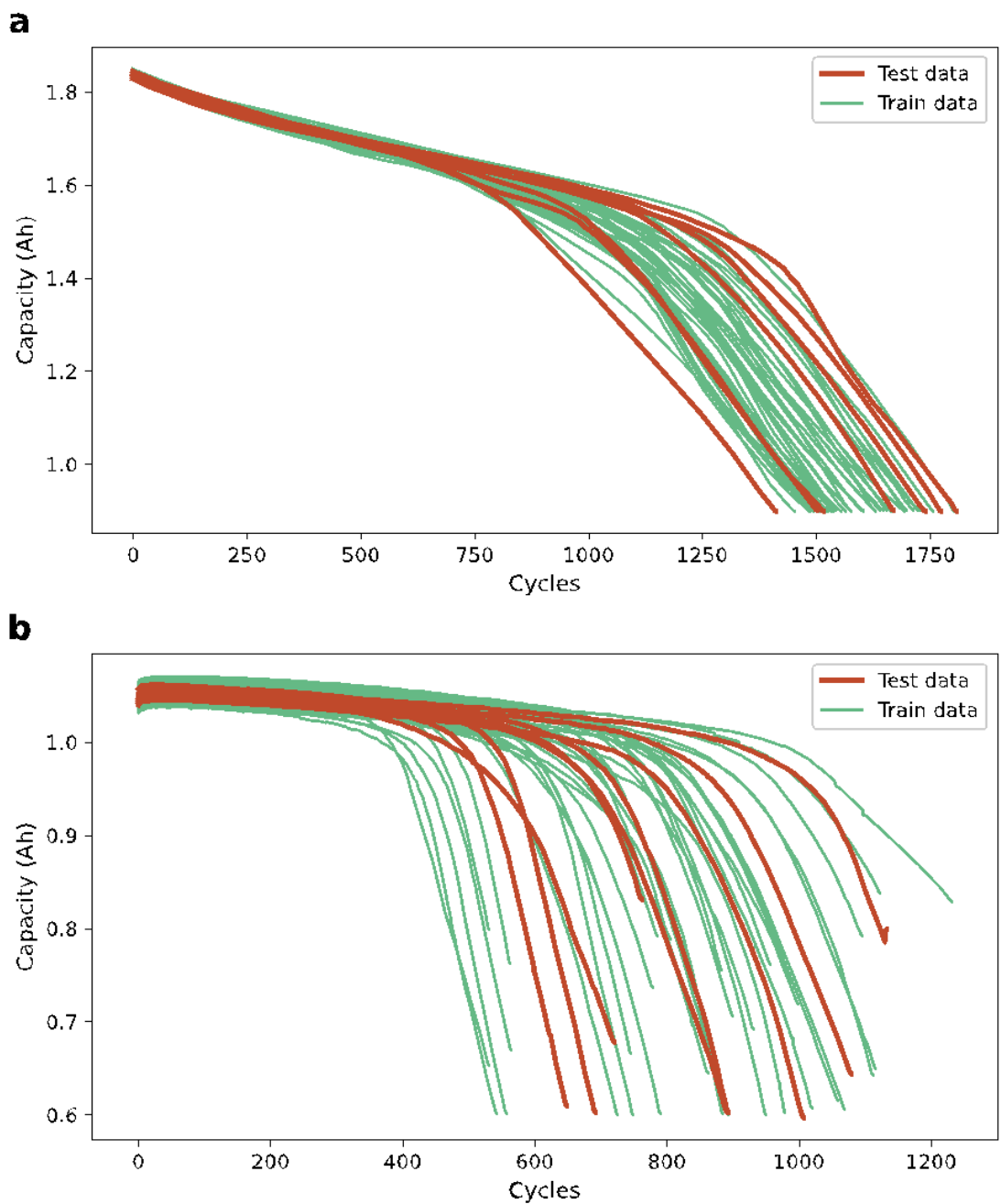

**Figure S1.** Training and validation/test batches of cells for the **a,** RWTH-ISEA and **b,** Stanford/MIT datasets. The cells were selected randomly.

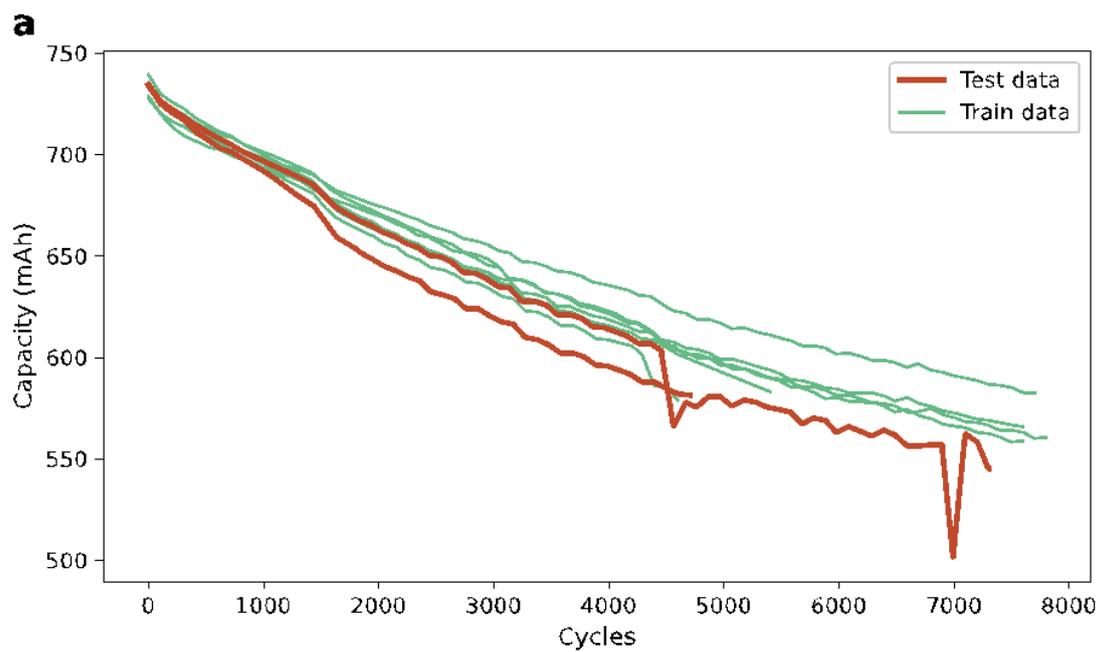

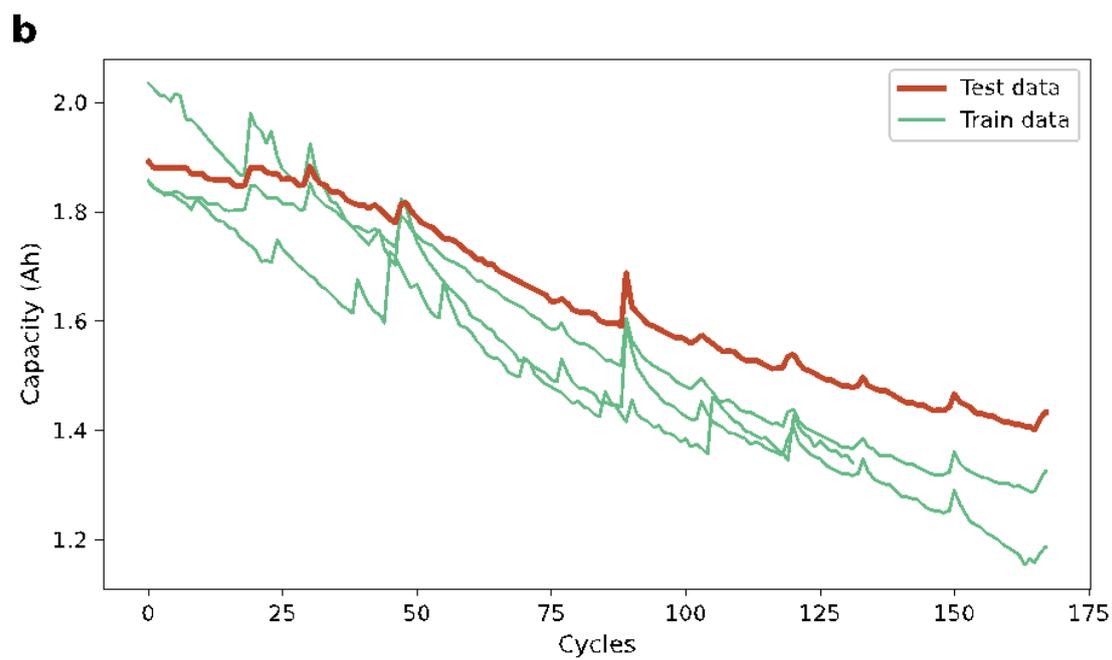

**Figure S2.** Training and validation (test) batches of cells for the **a**, Oxford and **b,** NASA datasets. The cells were selected randomly.

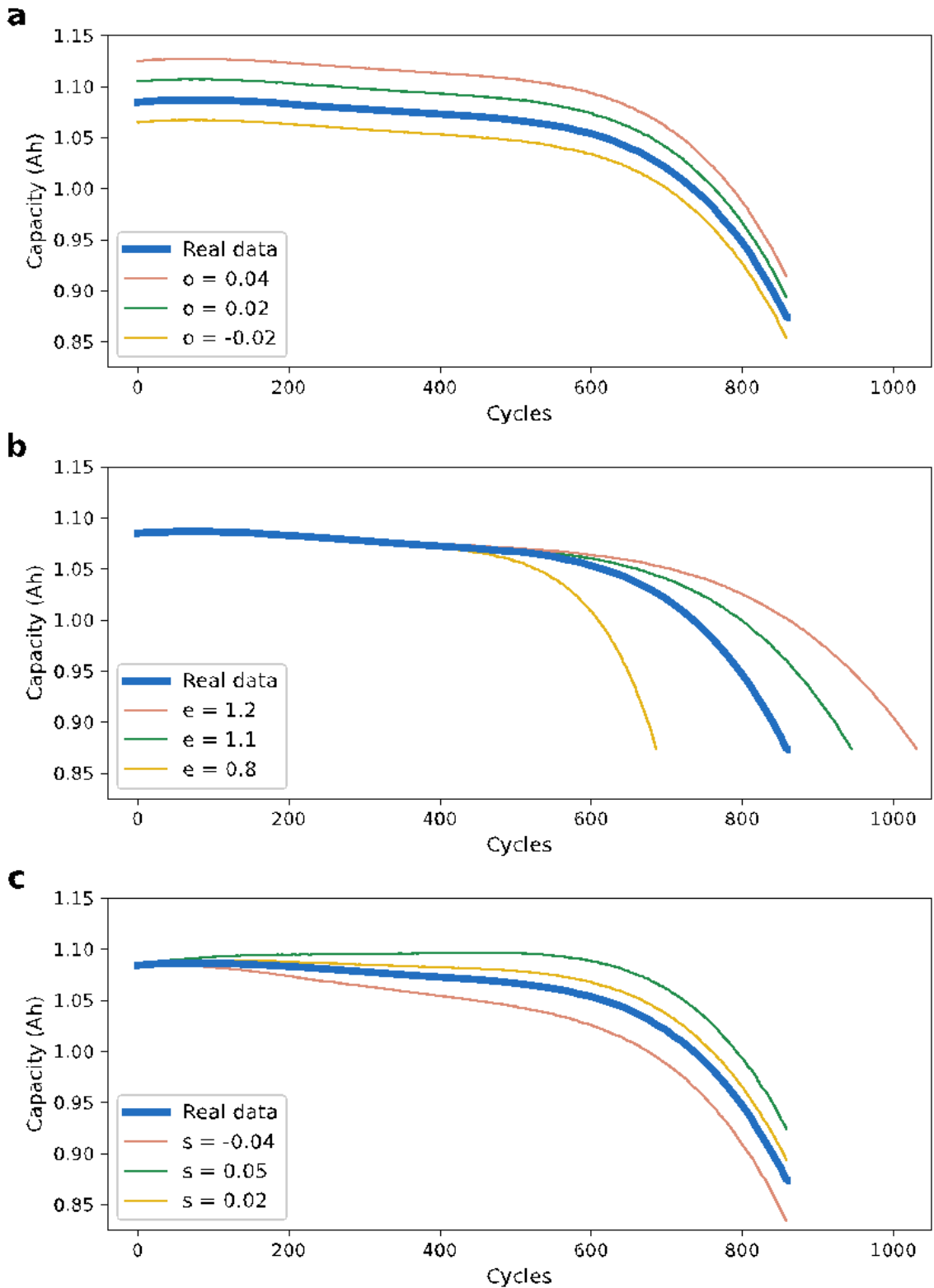

**Figure S3.** Quantitative variations of synthetic data based on the 3 parameters – (a) offset 'o', (b) elongation 'e' and (c) slope 's'. The offset parameter shifts the degradation curve in the vertical plane. The elongation parameter stretches or contracts the degradation curve, thereby creating battery degradation curves with different end-of-lives. The slope parameter accounts for the variation in the initial slopes of the degradation curves. It is evident that geometrical modifications can be performed to any extent and hence, there must be well-defined limits for the parameters to generate synthetic curves that represent the real dataset.

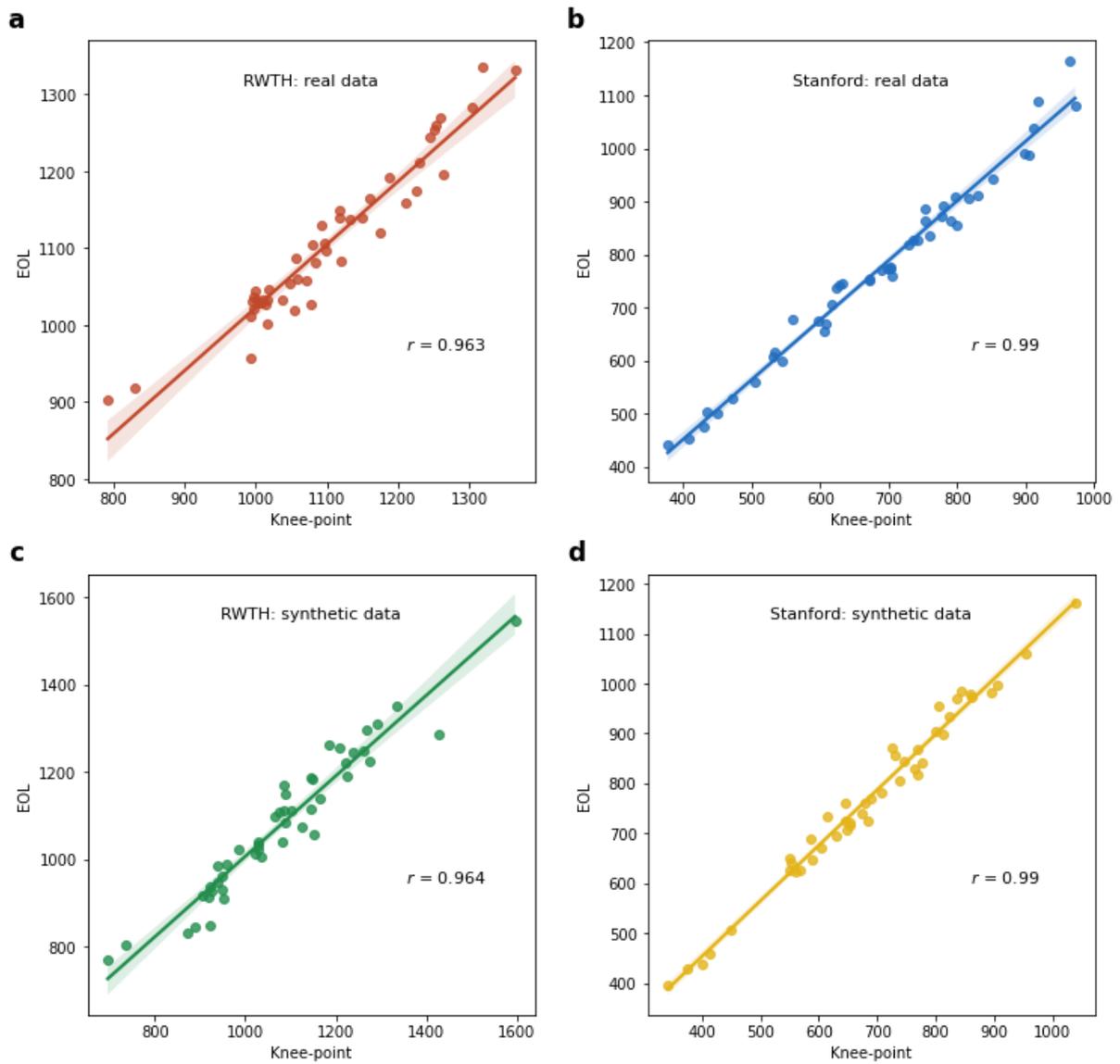

**Figure S4.** Regression plots showing correlations between the EOL and knee-points of the real capacity degradation data (a, b) and of the corresponding sets of randomly generated synthetic data (c, d) for the two datasets. The real data and respective randomly generated synthetic data have similar Pearson correlation coefficients, thereby affirming our approach to synthetic data generation. For the purpose of this figure, the numbers of synthetic data generated are equal to the numbers of real data in the respective datasets.

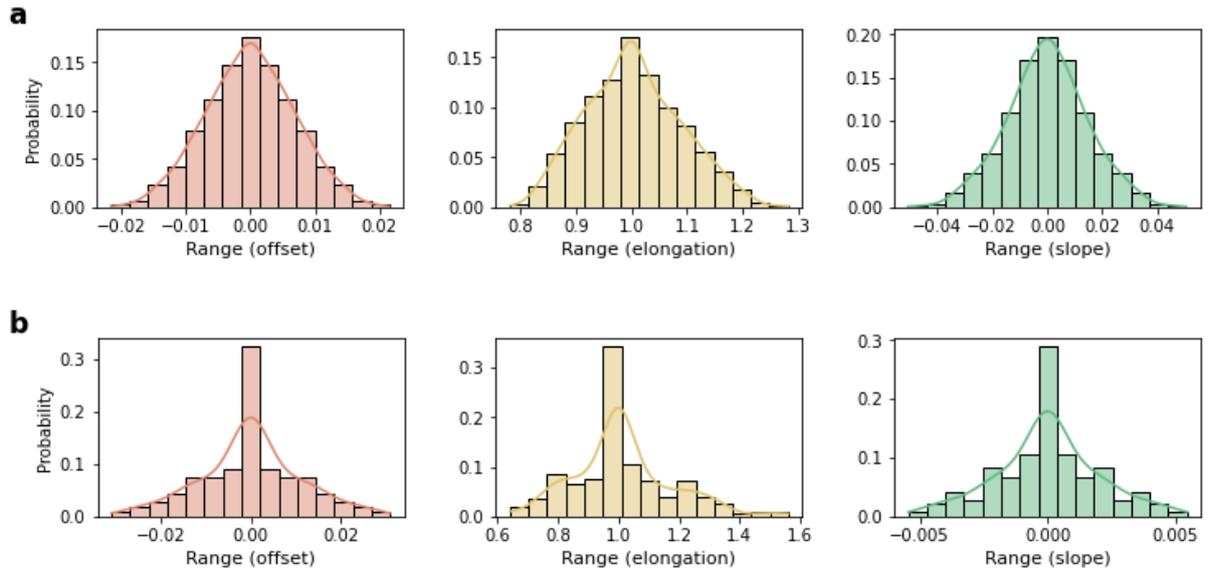

**Figure S5.** SDG function parameter distribution for the **a**, RWTH and **b**, Stanford datasets.

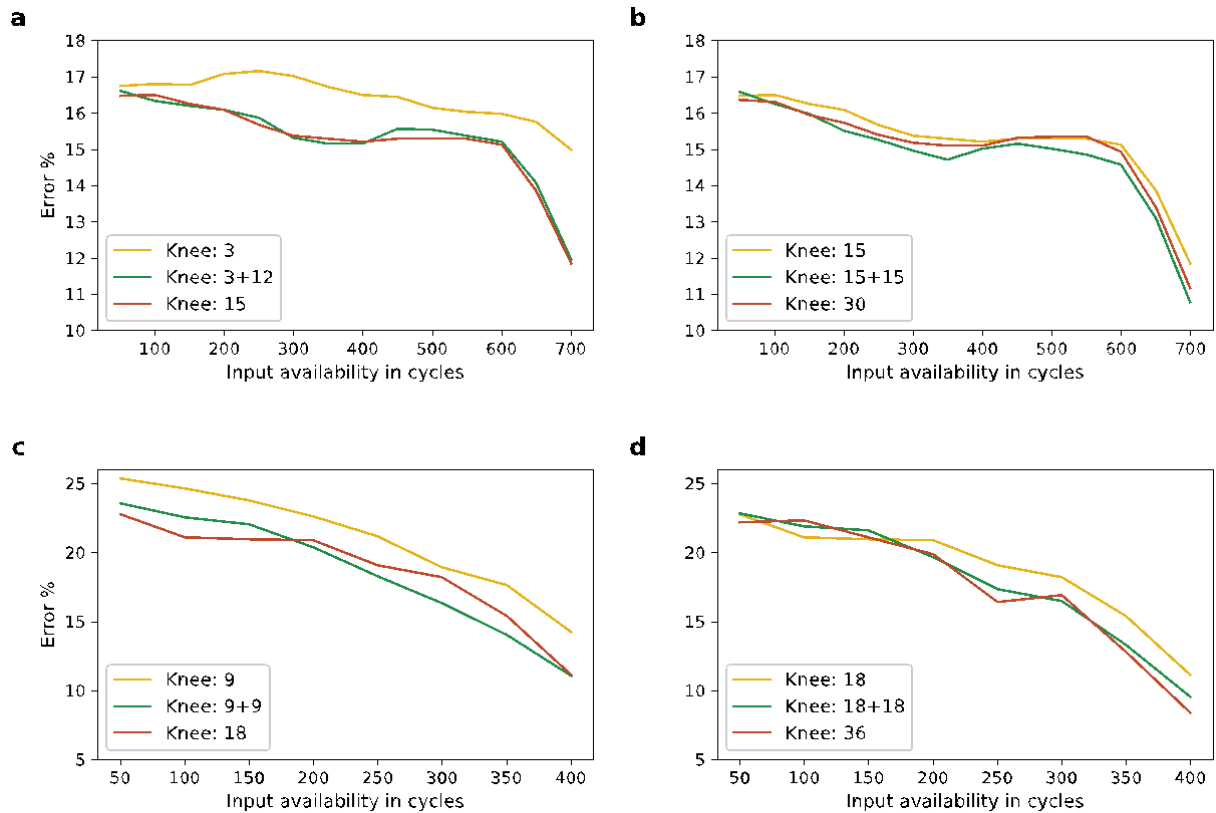

**Figure S6.** Performance of the CNN model upon replacement of real data by synthetic data for knee-point prediction for the RWTH dataset (a, b) and the Stanford dataset (c, d). The number of real degradation curves is denoted by a single integer 'n'. The numbers of real and synthetic degradation curves respectively within the training data are denoted by 'n+m'. It can be observed that the errors obtained upon replacement by synthetic data are similar to purely real data.

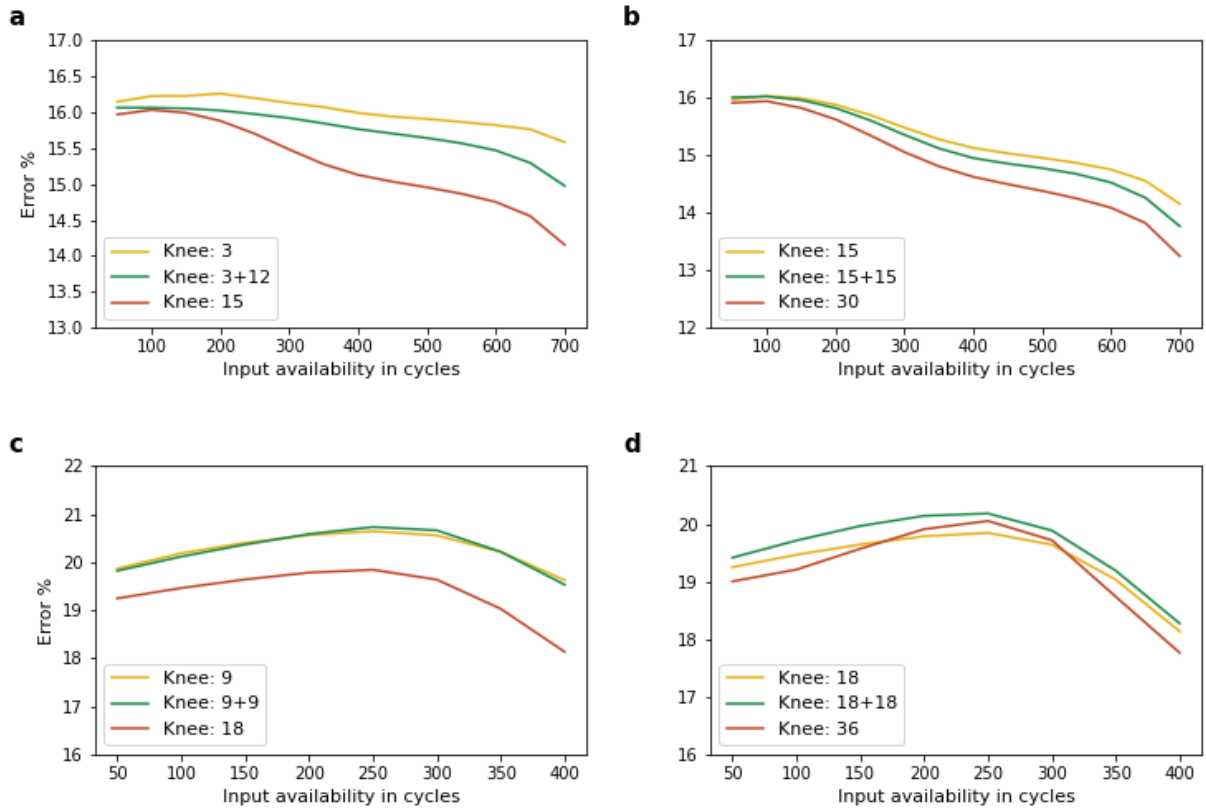

**Figure S7.** Performance of the GPR model upon replacement of real data by synthetic data for knee-point prediction for the RWTH dataset (a, b) and the Stanford dataset (c, d). The number of real degradation curves is denoted by a single integer 'n'. The numbers of real and synthetic degradation curves respectively within the training data are denoted by 'n+m'. It can be observed that the errors obtained upon replacement by synthetic data are similar to purely real data.

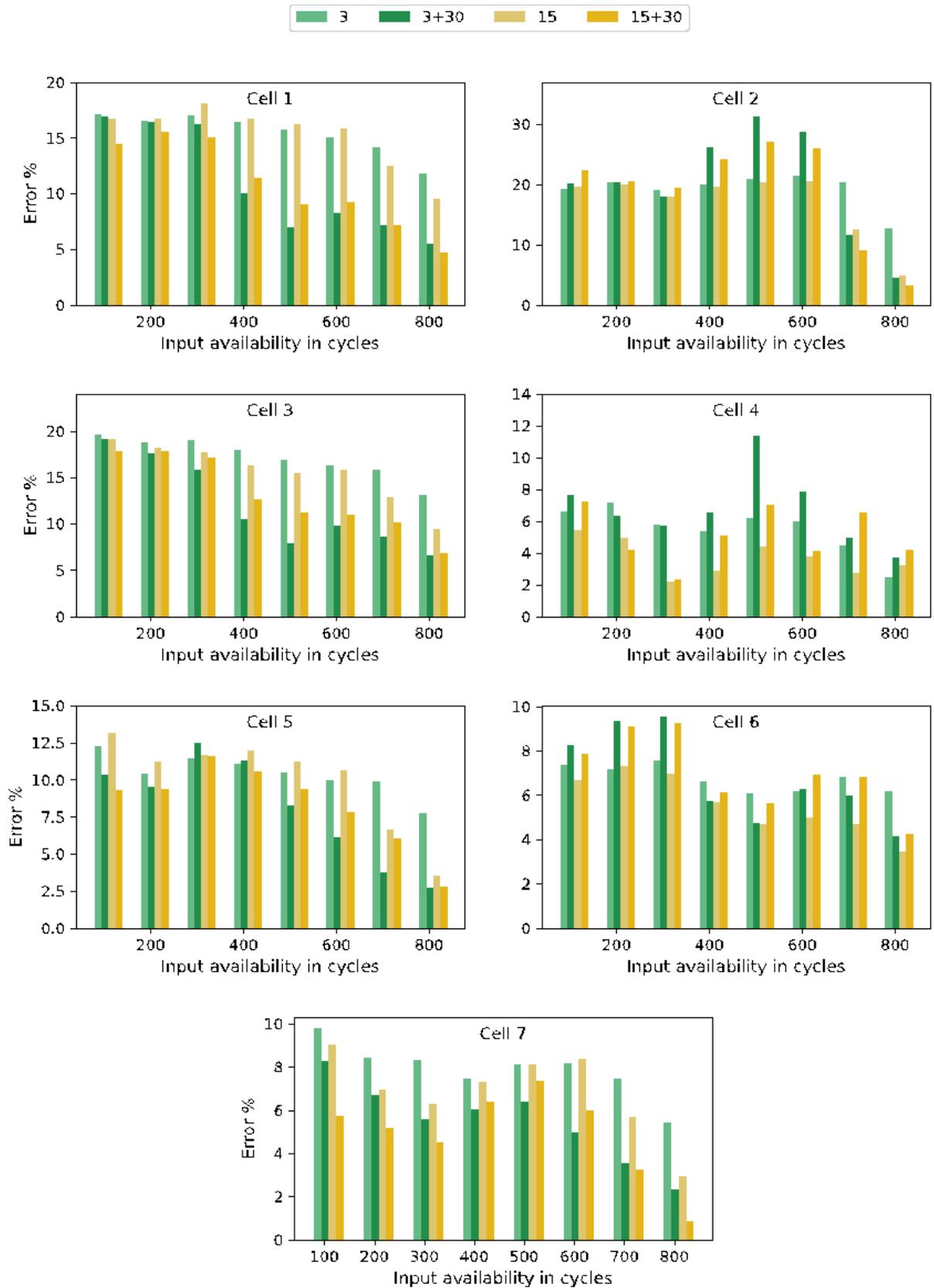

**Figure S9.** Mean errors of the CNN model in EOL prediction of individual test cells of the RWTH dataset. Model performance with real cells and combination of real and synthetic cells are labeled as 'n' and 'n+m' respectively.

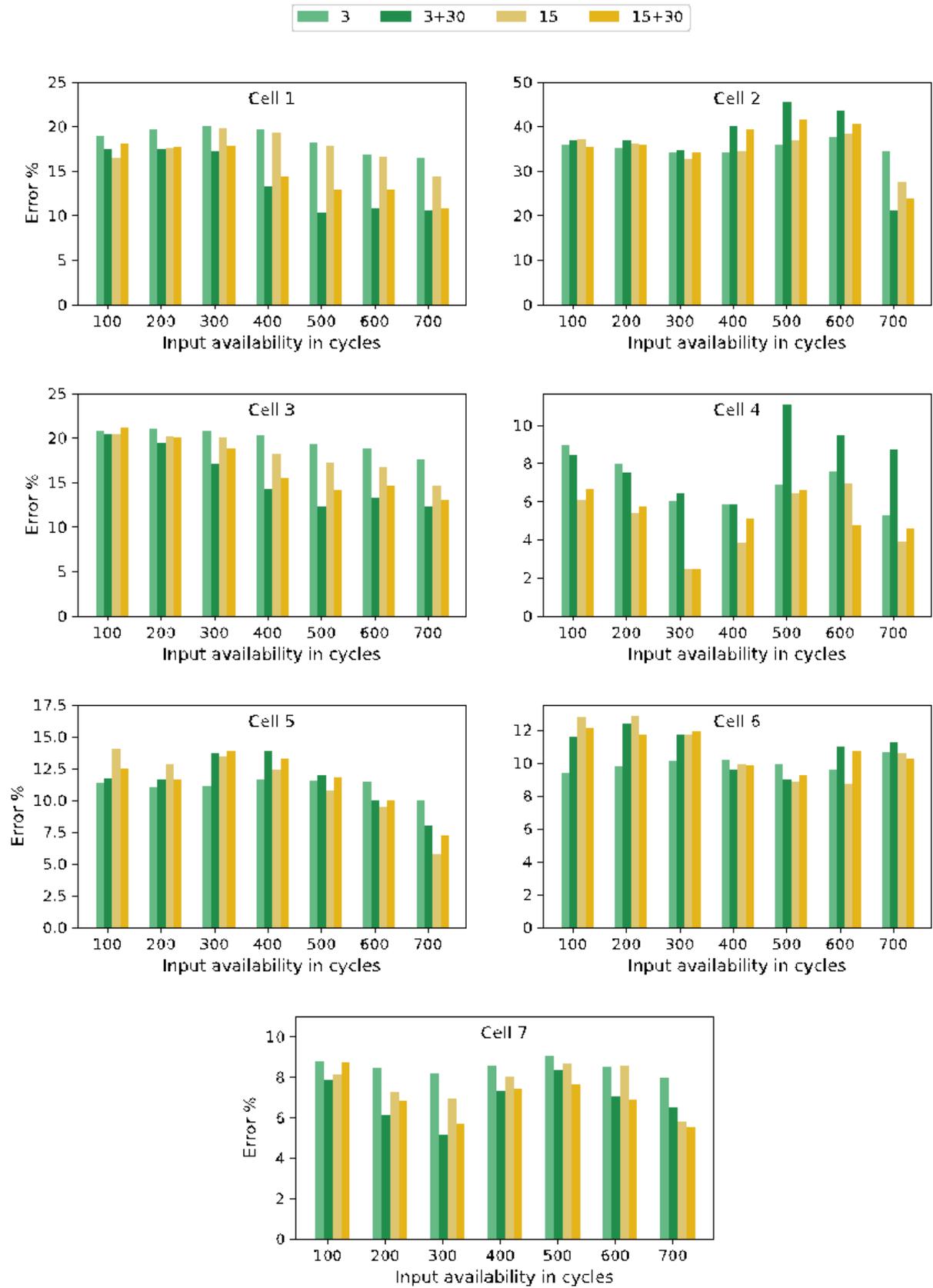

**Figure S10.** Mean errors of the CNN model in knee-point prediction of individual test cells of the RWTH dataset. Model performance with real cells and combination of real and synthetic cells are labeled as 'n' and 'n+m' respectively.

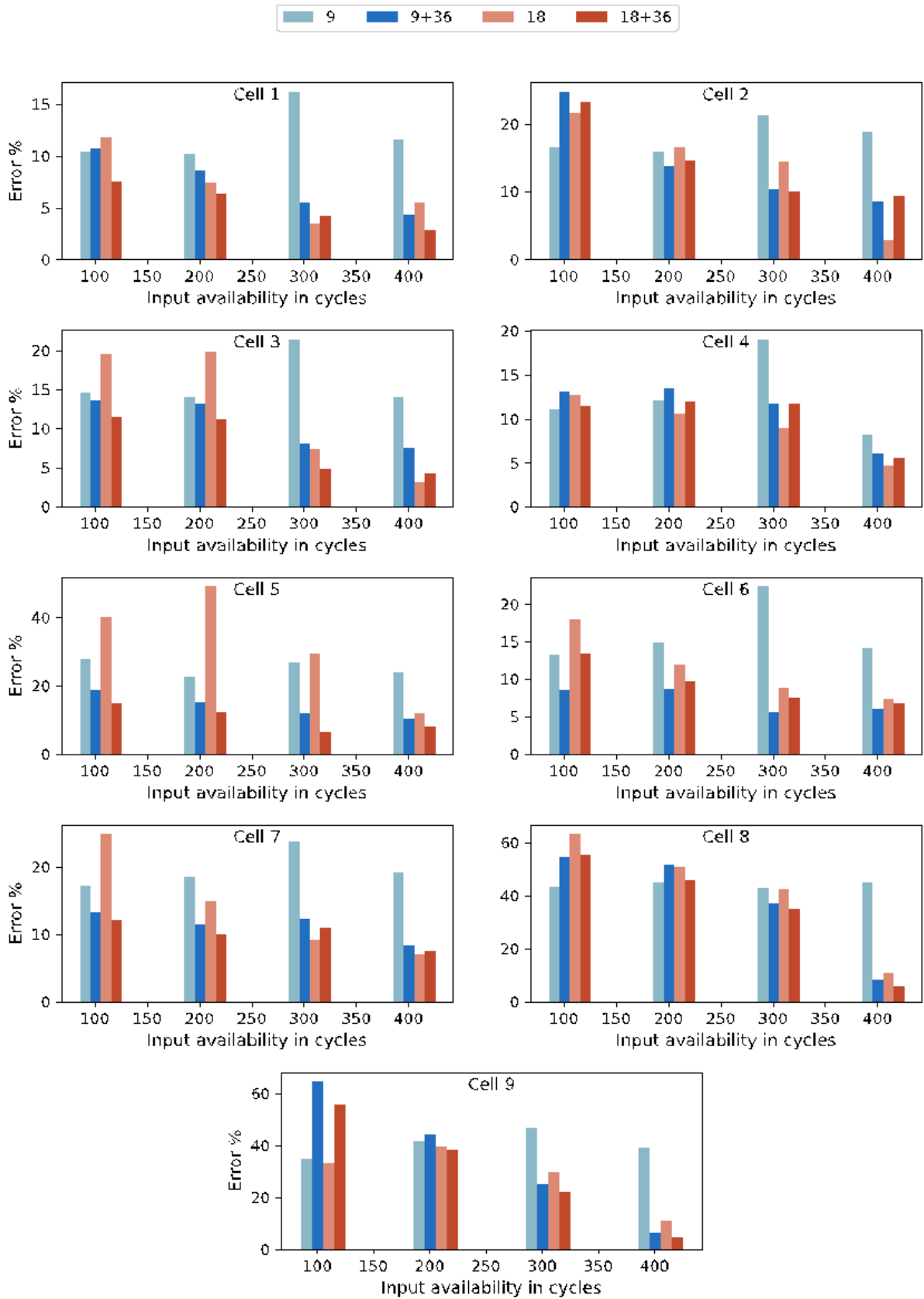

**Figure S11.** Mean errors of the CNN model in EOL prediction of individual test cells of the Stanford dataset. Model performance with real cells and combination of real and synthetic cells are labeled as 'n' and 'n+m' respectively.

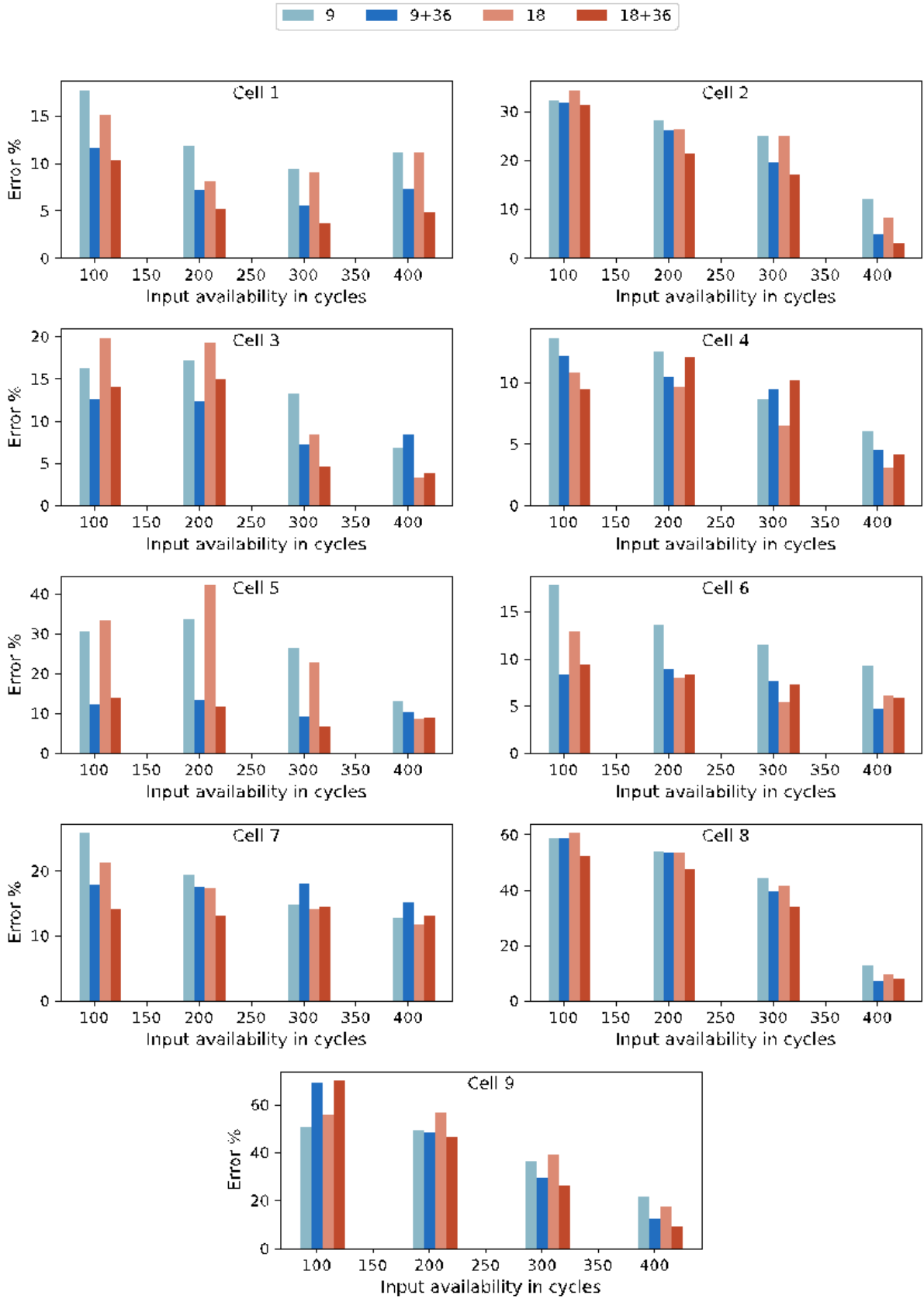

**Figure S12.** Mean errors of the CNN model in knee-point prediction of individual test cells of the Stanford dataset. Model performance with real cells and combination of real and synthetic cells are labeled as 'n' and 'n+m' respectively.

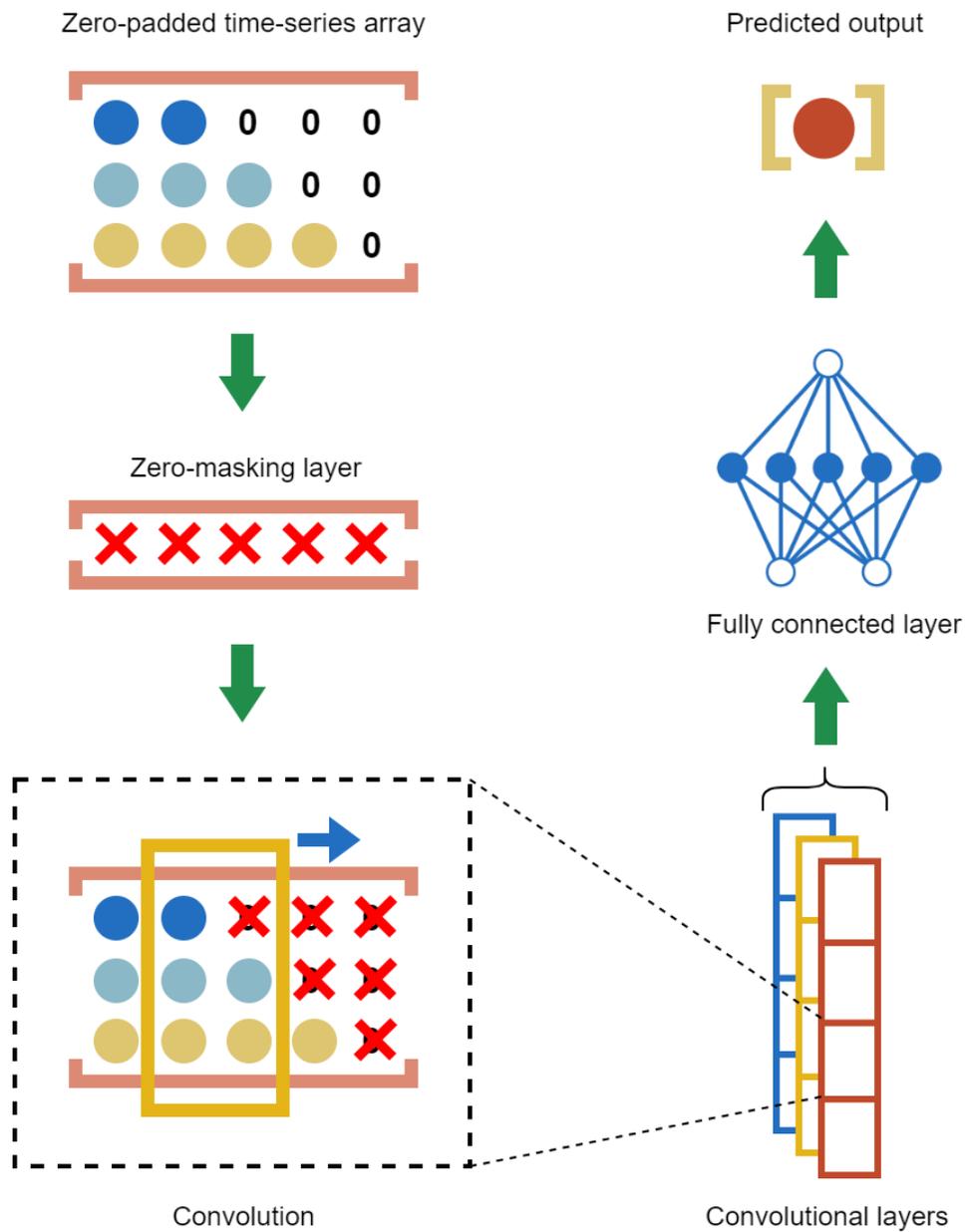

**Figure S13.** The framework and architecture of the CNN model used in the simulations. The predicted output can either be the knee-point or the EOL, or any other characteristic point that needs to be predicted.

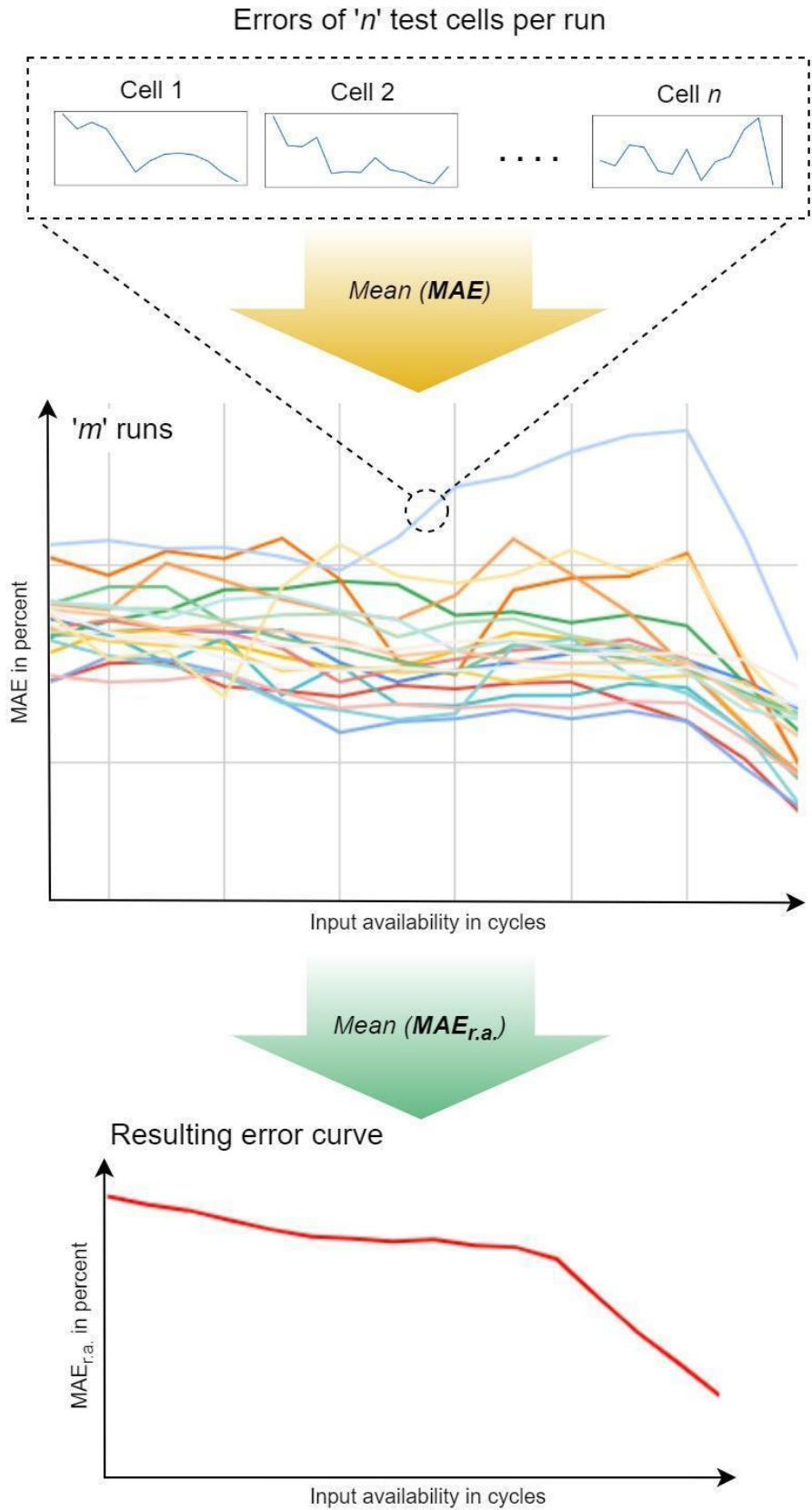

**Figure S14.** Method of calculation of mean error curves, starting at the cell-level errors to the result of averaging run-level error curves. The final curve is shown as errors in the Results section of the paper.

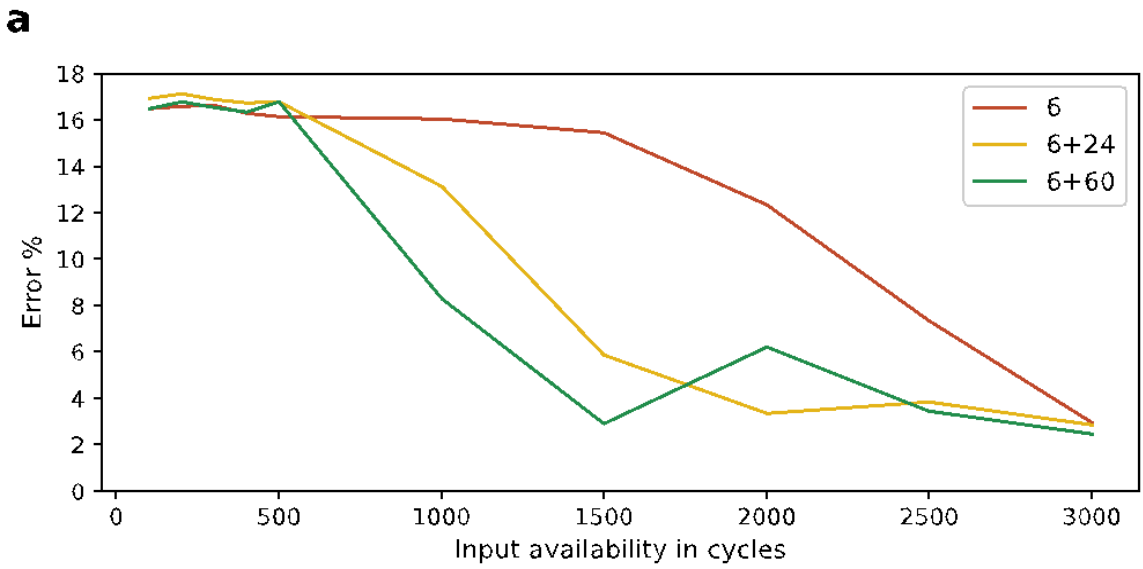

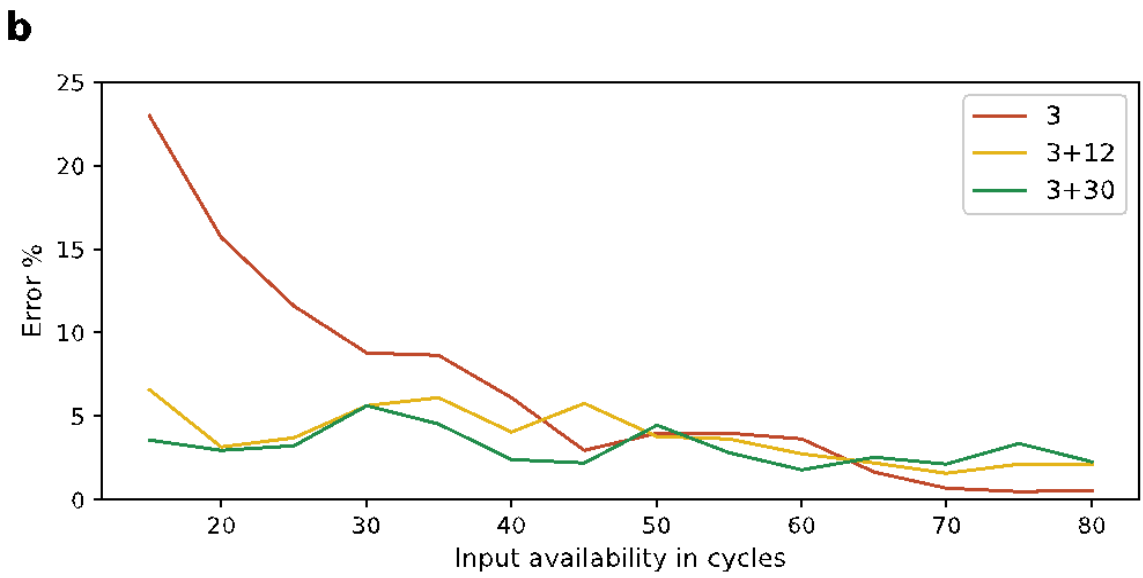

**Figure S15.** Prediction errors of the CNN model for the EOL points in the test cells of the **a,** Oxford and **b,** NASA datasets. Model performance with real cells and combination of real and synthetic cells are labelled as 'n' and 'n+m' respectively.

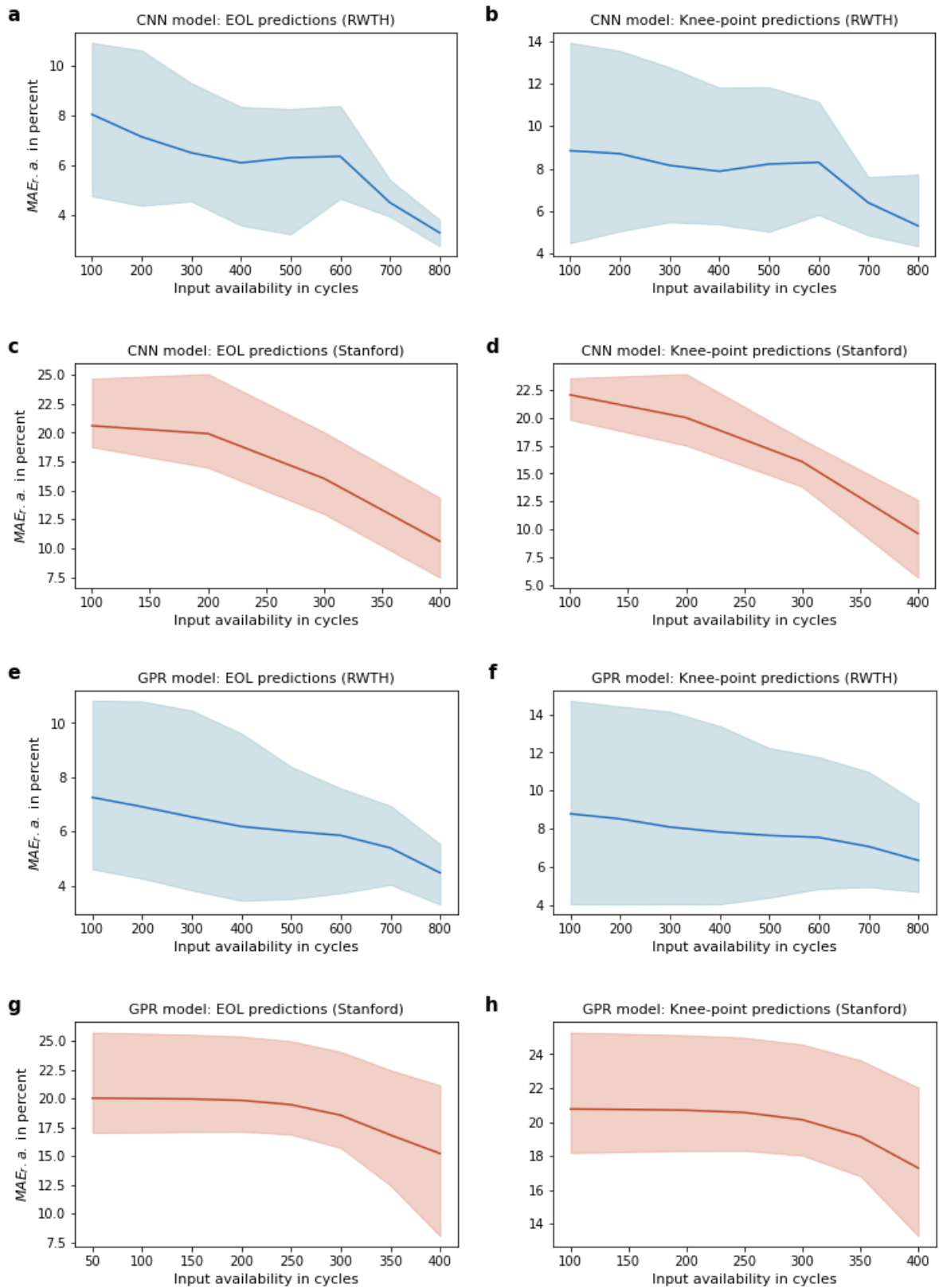

**Figure S16.** K-fold cross validation errors of the CNN and GPR models for the EOL and knee-points for the RWTH and Stanford datasets. The value of 'k' for the RWTH and Stanford datasets used were 6 and 5 respectively. The gray-shaded regions show the errors between the minimum and maximum error-boundaries, whereas the black lines show the mean errors at every cycle number.

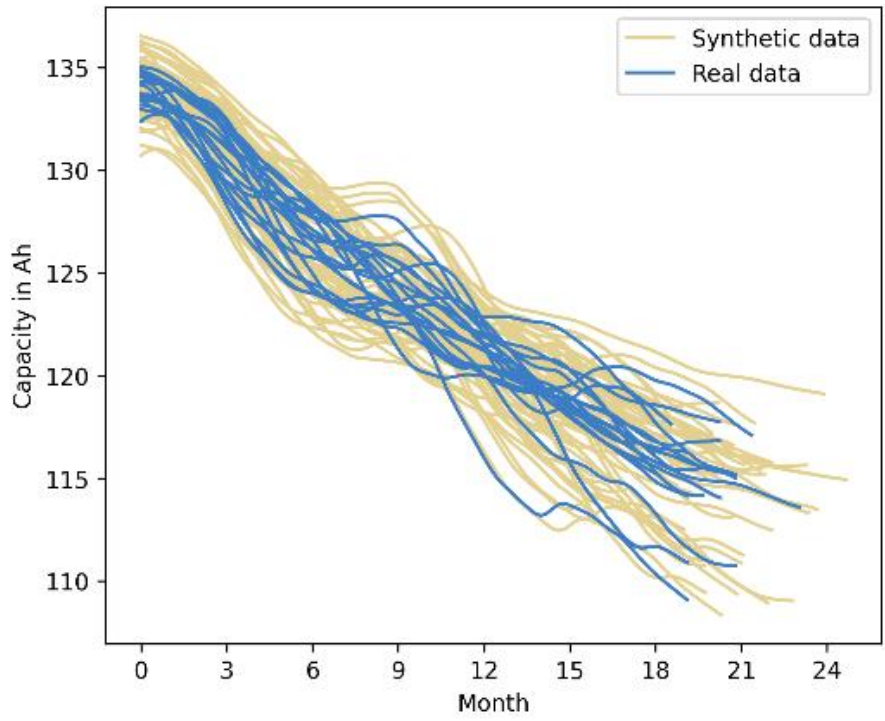

**Figure S17.** Field datasets from 20 electric vehicles, along with their sample synthetic curves.

Supplementary tables:

| Dataset | Indices | Knee-point | EOL |
|---|---|---|---|
| RWTH-ISEA dataset | Cell 1 | 1305 | 1336 |
| | Cell 2 | 792 | 941 |
| | Cell 3 | 1365 | 1402 |
| | Cell 4 | 1008 | 1072 |
| | Cell 5 | 1251 | 1306 |
| | Cell 6 | 1212 | 1223 |
| | Cell 7 | 1188 | 1240 |
| Stanford/MIT dataset | Cell 1 | 629 | 742 |
| | Cell 2 | 561 | 677 |
| | Cell 3 | 777 | 872 |
| | Cell 4 | 779 | 891 |
| | Cell 5 | 690 | 771 |
| | Cell 6 | 743 | 827 |
| | Cell 7 | 618 | 706 |
| | Cell 8 | 472 | 529 |
| | Cell 9 | 409 | 453 |
| Oxford dataset | Cell 1 | - | 4364 |
| | Cell 2 | - | 3474 |
| NASA dataset | Cell 1 | - | 83 |

**Table S1.** Values of knee-points and EOL80 (in cycles) of the validation batches of cells of the used laboratory datasets.

Supplementary notes:
**Note S1:**

Synthetic data generated through the method described in the paper is capable of saving significant effort for aging and generation of real data. To understand the degree of effort and cost savings, we refer to the results displayed in Figure 5 under the subsection "Partial replacement of real data with synthetic data".

Taking into account Figures 5 (a, e), the model performance by training on real data of 15 cells is similar to the training set of 3 real cells and 12 synthetic curves generated using the 3 real cells. The approximate savings in cell-testing effort in this case is equivalent to aging of (15 - 3) cells = 12 cells.

$$Effort\ Savings\ = \frac{15-3}{15} * 100\% \ = 80\%$$

Similarly, in Figures 5 (b, f), the model performance by training on real data of 30 cells is similar to the training set of 15 real cells and 15 synthetic curves. The approximate effort savings in this case is equivalent to aging of (30 - 15) cells = 15 cells.

$$Effort\ Savings\ = \frac{30-15}{30} * 100\% \ = 50\%$$

As a reminder, the process of synthetic data generation is random. Additionally, optimization of the prediction model parameters is not the focus of the work of this paper. It is expected that the model performance would increase by tuning the model parameters as well as the synthetic data generation process, which can further increase effort and cost savings.